\documentclass[%
 reprint,
 amsmath,amssymb,
 aps,
]{revtex4-2}

\usepackage{epstopdf}
\usepackage{graphicx}
\usepackage{dcolumn}
\usepackage{bm}

\usepackage{color}

\begin{document}

\preprint{APS/123-QED}

\title{Atomic-scale on-demand photon polarization manipulation with high-efficiency for integrated photonic chips}

\author{Yunning Lu$^{1,2}$}
\author{Zeyang Liao$^{1}$}
\thanks{liaozy7@mail.sysu.edu.cn}
\author{Xue-Hua Wang$^{1}$}%
 \thanks{wangxueh@mail.sysu.edu.cn}
 \affiliation{$^{{\small 1}}$\textit{ State Key Laboratory of Optoelectronic Materials and Technologies, School of Physics, Sun Yat-sen University, Guangzhou 510275, China}\\
$^{{\small 2}}$\textit{School of Electrical and Information Engineering, Anhui University of Technology, Ma'anshan, 243002, China}}




\date{\today}

\begin{abstract}
In order to overcome the challenge of lacking polarization encoding in integrated quantum photonic circuits, we propose a scheme to realize arbitrary polarization manipulation of a single photon by integrating a single quantum emitter in a photonic waveguide. In our scheme, one transition path of the three-level emitter is designed to simultaneously couples with two orthogonal polarization degenerate modes in the waveguide with adjustable coupling strengths, and the other transition path of the three-level emitter is driven by an external coherent field. The proposed polarization converter has several advantages, including arbitrary polarization conversion for any input polarization, tunable working frequency, excellent anti-dissipation ability with high conversion efficiency, and atomic-scale size. Our work provides an effective solution to enable the polarization encoding of photons which can be applied in the integrated quantum photonic circuits, and will boost quantum photonic chip.
\end{abstract}

\maketitle



Since photons have large degrees of freedom, long coherence time even at the room temperature and ultrafast transmission speed, they are ideal carriers of quantum information and are often used as quantum bits~\cite{Kok2007,Ladd2010,Llewellyn2020,Hochrainer2022}. Quantum advantages have been experimentally demonstrated using linear optical system \cite{HanSenZhong2021,Deng2023}, but the setup is usually very bulky. In recent years, integrated quantum photonic circuits (IQPC), in which the optical elements are integrated in a planar chip, have attracted extensive attentions due to its high stability, high scalability, high miniaturization, and high mobility \cite{JWang2020,Bogaerts2020,Elshaari2020,Arrazola2021,Vigliar2021,Dai2022,Madsen2022}. In contrast to the bulky linear optical systems where photon polarizations are often used as encoding quantum bits, the qubits in the IQPCs are usually encoded in the photon paths because there has been a lack of the solutions to on-demand arbitrary photon polarization manipulation on the chip \cite{Silverstone2013,WangJW2018,Politi2008,ZhangM2021,Morozko2023}. The challenge originates from the fact that the typical elements for polarization conversion, such as natural birefringent materials, Faraday magneto-optical rotator, and chiral metamaterials are usually too bulky to be integrated \cite{Born1999,Balos2020,Suresh2020,FCosta2020,SSui2016,JKGansel2009,Zhang2022,Chen2023}.


Photonic waveguide is a basic element in IQPC as the photon line to connect other photonic elements. By integrating quantum emitters into the waveguides and manipulate their coupling, known-as waveguide quantum electrodynamics, which allows to control the photonic degrees of freedom more conveniently and transfer the quantum information between distant nodes, is booming \cite{HuaixiuZheng2013,Roy2017,Liao2018,Li2018,Yang2022,Liao2010,Zanner2022,Liao2020,Kannan2023,Lodahl2015,LZhou2013,DEChang2018,Pennetta2022,Bag2023,Sheremet2023}. It has been shown that frequency and special polarization conversions can be realized by coupling a three-level $\Lambda$-type emitter with two different waveguide modes \cite{Pinotsi2008,Bradford2012PRL,Koshino2013PRL,Ballestero2016,TSTsoi2009,ZYZhang2013,Rosenblum2017,Pivovarov2021PRA,Wang2014,YLu2022,Shomroni2014,MTCheng2018,FQYu2019,Du2021}. To encode quantum information into photon polarization it is necessary to realize arbitrary rotation of polarization and generate arbitrary superposition of polarization states. However, up to now, no scheme can realize arbitrary polarization rotation of photons in the IQPC. 


In this work, we propose a scheme of integrating a single $\Lambda$-type three-level emitter in a semi-infinite rectangular waveguide to realize arbitrary polarization rotation of single photons where one atomic transition simultaneously couples with two orthogonal polarization degenerate modes (${\rm TE_{01}}$ and ${\rm TE_{10}}$), and the other atomic transition is driven by an adjustable external field. Our scheme provides multi-dimensional control to transform an input photon with any polarization into an output photon with other arbitrary polarization which is vital for polarization encoding of photons on chips. Our scheme has several other advantages. First, the working frequency is tunable. Second, it possesses excellent anti-dissipation ability with high conversion efficiency due to the effect of electromagnetically induced transparency (EIT). Finally, the size of our polarization converter is of atomic scale, in vast favor of on-chip integration.


\begin{figure}[!ht]
\centering
\includegraphics[width=0.99\columnwidth]{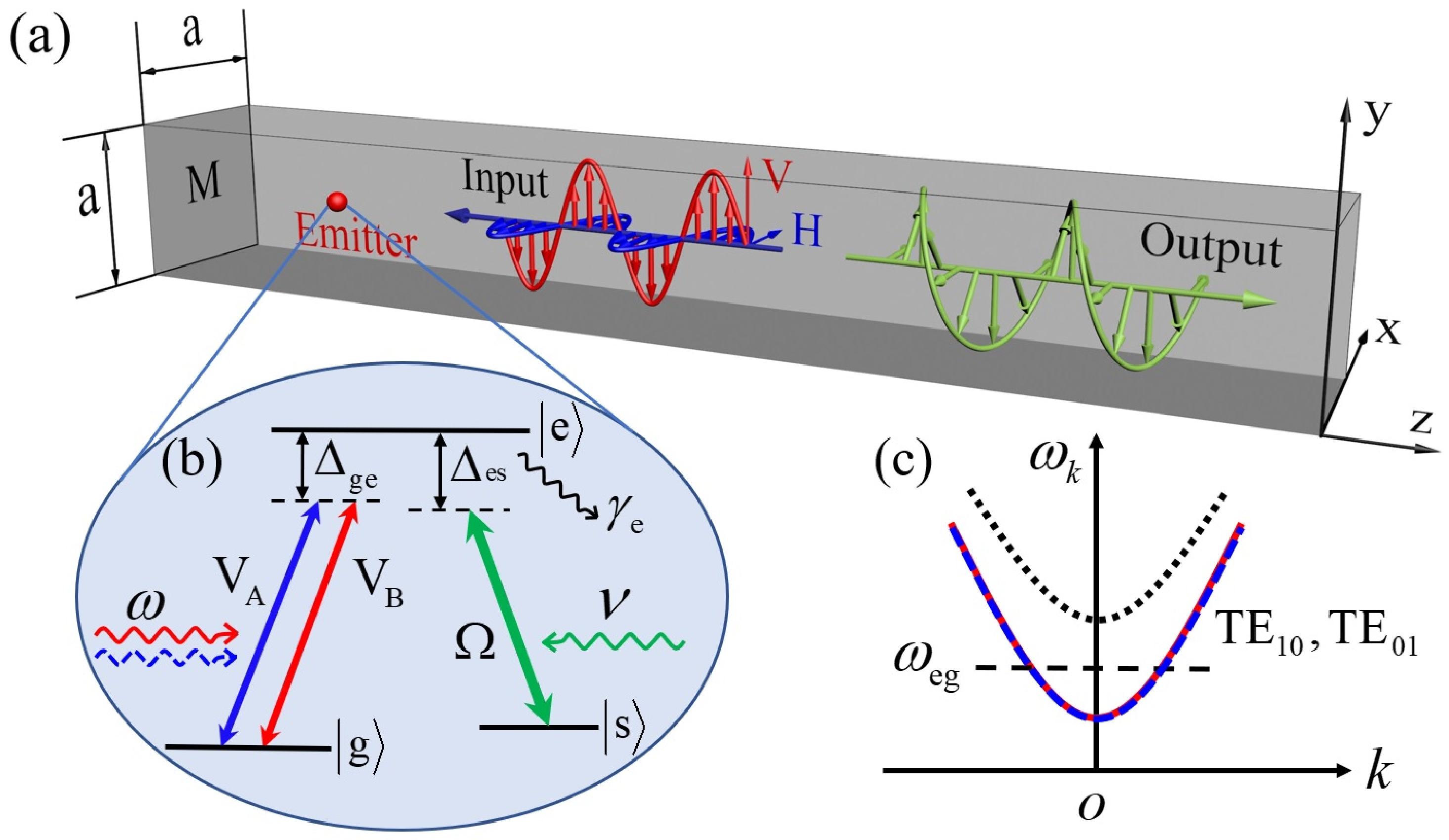}
\caption{(a) Schematic diagram of the integrated photon polarization converter. 
(b) Energy levels of the quantum emitter. 
(c) The dispersion relations of the degenerate modes ${\rm TE_{01}}$ and ${\rm TE_{10}}$ (blue solid curve) and two higher-energy modes (red dashed curve and black dotted curve) of the waveguide. 
$\omega_2$ is the energy of state $|{\rm e}\rangle$.
\label{fig:Model}}
\end{figure}


The model we consider is shown in Fig.~\ref{fig:Model}(a) where a single $\Lambda-$type quantum emitter is coupled to a semi-infinite square waveguide with cross section $a\times a$~\cite{JunhongAn2021,JunhongAn2024,ZhouLan2014,JunhongAn2017,JieyuYou2018,JinfengHuang2013}. In the square waveguide, ${\rm TE_{mn}}$ and ${\rm TE_{nm}}$ modes are degenerate. 
The emitter sitting at the center of the waveguide has three energy states denoted as $|{\rm g}\rangle$, $|{\rm e}\rangle$, and $|{\rm s}\rangle$, with energy $\omega_{\rm g}$, $\omega_{\rm e}$, and $\omega_{\rm s}$ (we set $\hbar=1$), respectively (Fig. 1(b)). 
 The states $|{\rm g}\rangle$ and $|{\rm s}\rangle$ are assumed to be metastable, and the state $|{\rm e}\rangle$ may dissipate energy into non-waveguide modes at rate $\gamma_{\rm e}$.  
Here we consider that ${\rm TE_{01}}$ mode (horizontally polarized, denoted as A mode) and ${\rm TE_{10}}$ mode (vertically polarized, denoted as B mode) can couple to the emitter transition $|{\rm g}\rangle \to |{\rm e}\rangle$ with coupling strengths $V_{\rm A}$ and $V_{\rm B}$, respectively (Fig. 1(c)). 
Other waveguide modes are large detuned from the $|{\rm g}\rangle \to |{\rm e}\rangle$ transition, and can be neglected. An additional external control field with angular frequency $\nu$ is applied to drive the $|{\rm e}\rangle \to |{\rm s}\rangle$ transition with coupling strength $\Omega$ and detuning $\Delta_{\rm es}=\omega_{\rm es}-\nu$.
When a left-propagating photon enters the waveguide, it can be reflected by the waveguide end, and become a symmetric superposition of the left- and right-propagating modes and the emitter is located at the antinode position.

The Hamiltonian of the system is given by 
$H = H_{\rm F} + H_{\rm A} + H_{\rm int}$
where 
$H_{\rm F} 
= \int{\rm d}z [a_{\rm R}^{\dagger}(z) (\omega_0 -iv_{\rm g}\frac{\rm d}{{\rm d}z}) a_{\rm R}(z)
+a_{\rm L}^{\dagger}(z) (\omega_0 +iv_{\rm g}\frac{\rm d}{{\rm d}z})  a_{\rm L}(z)]
+ \int{\rm d}z [b_{\rm R}^{\dagger}(z) (\omega_0 -iv_{\rm g}\frac{\rm d}{{\rm d}z}) b_{\rm R}(z)
+b_{\rm L}^{\dagger}(z) (\omega_0 +iv_{\rm g}\frac{\rm d}{{\rm d}z})  b_{\rm L}(z)]$ is the Hamiltonian of waveguide field under the linearization approximation with $a_{\rm R}$ ($b_{\rm R}$) and $a_{\rm L}$ ($b_{\rm L}$) being the annihilation operators of the right- and left-propagating A- (B-) mode photons, respectively \cite{Shen2005,ZeyangLiao2016}. The photon frequency $\omega=\omega_0+v_{\rm g}k$ where $\omega_0$ is a reference frequency around the photon frequency and $v_{\rm g}$ is the group velocity at $\omega_0$. 
$H_{\rm A}$ is the effective emitter Hamiltonian including the dissipation 
$H_{\rm A} =\omega_{\rm g}|{\rm g}\rangle \langle {\rm g}| + (\omega_{\rm e}-i\gamma_{\rm e}/2)|{\rm e}\rangle \langle {\rm e}| + (\omega_{\rm e}-\Delta_{\rm es})|{\rm s}\rangle \langle {\rm s}|
+ \Omega/2(|{\rm e}\rangle \langle {\rm s}|+{\rm H.c.})$
in the rotating frame with respect to driving frequency $\nu$ and $\omega_{g}$ can be set to zero. 
$H_{\rm int} =\sum_{{\rm p}={\rm R,L}}\int{\rm d}z \delta(z-z_0)[V_{\rm A} a_{\rm p}^{\dagger}(z)|{\rm g}\rangle \langle {\rm e}| + V_{\rm B} b_{\rm p}^{\dagger}(z)|{\rm g}\rangle \langle {\rm e}| + {\rm H.c.}]$ is the interaction Hamiltonian 
where $z_0$ is the $z$-coordinate of the emitter and $V_{\rm A,B}$ are the coupling strengths with two orthogonal modes. Assume that the transition dipole moment $\vec{\mu}_{\rm ge}$ is in the $x-y$ plane with $\theta$ being the angle between $\vec{\mu}_{\rm ge}$ and the $x$-axis which can be tuned by external electromagnetic field \cite{Buryakov-2023}, and we have  
$V_{\rm A}=-\sqrt{L}\mu_{\rm ge}E_{\rm A}\cos\theta/\hbar$ and $V_{\rm B}=-\sqrt{L}\mu_{\rm ge}E_{\rm B}\sin\theta/\hbar$.

Suppose that a left-propagating A-mode photon with angular frequency $\omega$ and wave vector $k$ is interacting with the emitter initially in the ground state, i.e.,  
$|\Psi^{(\rm A)}_{\rm L}\rangle = \int{\rm d}z e^{-ikz}/{\sqrt{2\pi}} a_{\rm L}^{\dagger}(z) |{\rm g},0\rangle$.
After scattering, the output state  
$|\Psi_{\rm out}\rangle=r_{\rm AA}(k)|\Psi^{(\rm A)}_{\rm R}\rangle+r_{\rm BA}(k)|\Psi^{(\rm B)}_{\rm R}\rangle $
where 
$|\Psi^{(\rm A)}_{\rm R}\rangle = \int{\rm d}z e^{ikz}/{\sqrt{2\pi}} a_{\rm R}^{\dagger}(z) |{\rm g},0\rangle$ and
$|\Psi^{(\rm B)}_{\rm R}\rangle = \int{\rm d}z e^{ikz}/{\sqrt{2\pi}} b_{\rm R}^{\dagger}(z) |{\rm g},0\rangle$
denote the right propagating A- and B- modes, respectively,
with coefficiencies \cite{sm}
\begin{eqnarray}
r_{\rm AA}(k) &=& \frac{-2\Gamma_{0}\text{cos}(2\theta)+\gamma_{\rm e}/2-i\alpha}
{2\Gamma_{0}+\gamma_{\rm e}/2-i\alpha},\label{eq:rAA}\\
r_{\rm BA}(k) &=& \frac{-2\Gamma_{0}\text{sin}(2\theta)}
{2\Gamma_{0}+\gamma_{\rm e}/2-i\alpha},\label{eq:rBA}
\end{eqnarray}
where $\alpha \equiv \Omega^2/[4(\Delta_{\rm ge}-\Delta_{\rm es})]-\Delta_{\rm ge}$ depends on the external controlling field and $\Delta_{\rm ge}=\omega_{e}-\omega$ is the detuning as shown in Fig. 2(b). 
$\Gamma_0=\mu_{\rm ge}^{2}\omega/(2\hbar\epsilon v_{\rm g}S_{\rm eff})$ with $\epsilon$ and $S_{\rm eff}$ being the dielectric constant and the effective cross section area, respectively.
Under two-photon resonance condition, i.e., $\Delta_{\rm ge}=\Delta_{\rm es}$,  we have $r_{\rm AA}=1$ and $r_{\rm BA}=0$ due to the  EIT effect.
Similarly, when a B-mode photon is injected, the output wavefunction $|\Psi_{\rm out}\rangle=r_{\rm AB}(k)|\Psi^{(\rm A)}_{\rm R}\rangle+r_{\rm BB}(k)|\Psi^{(\rm B)}_{\rm R}\rangle$ with coefficients 
\begin{eqnarray}
r_{\rm BB}(k) &=& \frac{2\Gamma_{0}\text{cos}(2\theta)+\gamma_{\rm e}/2-i\alpha}
{2\Gamma_{0}+\gamma_{\rm e}/2-i\alpha},\label{eq:rBB}\\
r_{\rm AB}(k) &=& \frac{-2\Gamma_{0}\text{sin}(2\theta)}
{2\Gamma_{0}+\gamma_{\rm e}/2-i\alpha}.\label{eq:rAB}
\end{eqnarray}
Again, $r_{\rm AB}=0$ and $r_{\rm BB}=1$ when $\Delta_{\rm ge}=\Delta_{\rm es}$.

For convenience, we denote a photon in the superposition of A- and B- mode with coefficients $C^{{\rm(A)}}$ and $C^{{\rm (B)}}$ by the vector $[C^{{\rm (A)}},C^{{\rm (B)}}]^{\rm T}$.
Thus, A- and B- mode photon are denoted by $|{\rm H}\rangle=[1,0]^{\rm T}$ and $|{\rm V}\rangle=[0,1]^{\rm T}$, respectively.
In general, if a single photon state
$|\Psi_{\rm in}\rangle=[C^{\rm (A)}_{\rm in},C^{\rm (B)}_{\rm in}]^{\rm T}$
is input and scattered by the emitter,
the output state $|\Psi_{\rm out}\rangle=[C^{\rm (A)}_{\rm out},C^{\rm (B)}_{\rm out}]^{\rm T}={\mathcal{S}}|\Psi_{\rm in}\rangle$, where ${\mathcal{S}}$ is the scattering matrix
\begin{eqnarray}
{\mathcal{S}}=
\left[
  \begin{array}{cc}
    r_{\rm AA}(k) & r_{\rm AB}(k) \\
    r_{\rm BA}(k) & r_{\rm BB}(k) \\
  \end{array}
\right].\label{eq:matrix}
\end{eqnarray}


\textit{Proof of arbitrary polarization conversions-}Consider an arbitrary input state $|\Psi_{\rm in}\rangle=[I_{\rm A} e^{i\xi_{\rm I}},I_{\rm B}]^{\rm T}$ and an arbitrary output state $|\Psi_{\rm out}\rangle=[O_{\rm A} e^{i(\xi_{\rm co}+\xi_{\rm O})},O_{\rm B}e^{i\xi_{\rm co}}]^{\rm T}$ where $0 \leq I_{\rm A},\ I_{\rm B},\ O_{\rm A},\ O_{\rm B}\leq 1$, $I_{\rm A}^2 + I_{\rm B}^2 = 1$ and $O_{\rm A}^2 + O_{\rm B}^2 = 1$.
The phases $\xi_{\rm I}$ and $\xi_{\rm O}$ satisfy $-\pi \leq \xi_{\rm I}, \xi_{\rm O} < \pi$, and 
$\xi_{\rm co}$ is the global phase of the output state which is irrelevant. In the case when the external dissipation $\gamma_{\rm e}$ is negligible \cite{Sheremet2023,Scarpelli2019}, by solving the equation $|\Psi_{\rm out}\rangle=\mathcal{S}|\Psi_{\rm in}\rangle$, we can obtain two equations shown in Eqs. (S61) and (S62) in \cite{sm} whose solutions are given by
\begin{eqnarray}
\alpha &=& \frac{\pm2\Gamma_0(I_{\rm A}I_{\rm B} \sin \xi_{\rm I} + O_{\rm A}O_{\rm B} \sin \xi_{\rm O})}
{\sqrt{(I_{\rm A}^2- O_{\rm A}^2)^2+(I_{\rm A}I_{\rm B} \cos \xi_{\rm I} - O_{\rm A}O_{\rm B} \cos \xi_{\rm O})^2}}, \nonumber \\ \\
\sin 2\theta &=& \frac{\pm(I_{\rm A}^2- O_{\rm A}^2)}
{\sqrt{(I_{\rm A}^2- O_{\rm A}^2)^2+(I_{\rm A}I_{\rm B} \cos \xi_{\rm I} - O_{\rm A}O_{\rm B} \cos \xi_{\rm O})^2}},\nonumber \\ \\
\cos 2\theta &=& \frac{\pm(-I_{\rm A}I_{\rm B} \cos \xi_{\rm I} + O_{\rm A}O_{\rm B} \cos \xi_{\rm O})}
{\sqrt{(I_{\rm A}^2- O_{\rm A}^2)^2+(I_{\rm A}I_{\rm B} \cos \xi_{\rm I} - O_{\rm A}O_{\rm B} \cos \xi_{\rm O})^2}}. \nonumber \\
\end{eqnarray}
Thus, by adjusting the external control field we can obtain the required values of $\theta$ and $\alpha$ which can convert the input state $|\Psi_{\rm in}\rangle$  to the desired output state $|\Psi_{\rm out}\rangle$. Several concrete examples are presented in the following. 

\textit{Conversion between a linearly polarized photon and a circularly polarized photon-}Assume that the emitter is isotropic or its transition dipole moment is along the angular bisector direction of $x$-axis and $y$-axis, i.e., $\theta=\pi/4$.
If the external driving field is adjusted to satisfy the condition $\alpha=\pm 2\Gamma_0$, i.e., $\Omega=\Omega_\pm\equiv 2\sqrt{(\Delta_{\rm ge}-\Delta_{\rm es})(\Delta_{\rm ge}\pm 2\Gamma_0)}$, the scattering matrix in Eq. (\ref{eq:matrix}) is reduced to
\begin{eqnarray}
{\mathcal{S}_\pm}
=
\frac{\mp i}{\sqrt{2}}
\left[
  \begin{array}{cc}
    e^{\pm i\pi/4} & e^{\mp i\pi/4} \\
    e^{\mp i\pi/4} & e^{\pm i\pi/4} \\
  \end{array}
\right].
\label{eq:S1}
\end{eqnarray}
When the incident photon is horizontally (vertically) polarized, the output photon $|\Psi_{\rm out}\rangle = [e^{i\pi/4},e^{-i\pi/4}]^{\rm T}/\sqrt{2}$ ($|\Psi_{\rm out}\rangle = [e^{-i\pi/4},e^{i\pi/4}]^{\rm T}/\sqrt{2}$) is right- (left-) handed circularly polarized if $\Omega=\Omega_{+}$, while it is left- (right-) handed circularly polarized if  $\Omega=\Omega_{-}$.
On the contrary, if the incident photon is right- or left-handed circularly polarized, they can be transformed to either horizontal or vertical polarization which depends on the control field as shown in Fig. 2(a).
Here, it should be noted that the right- (left-) handed circularly polarized input photon has the same vector expression with the left- (right-) handed circularly polarized output photon since the input and output photons have opposite propagation directions.

\textit{Arbitrary rotation of a linearly polarized photon-}
Considering the condition $\alpha=0$, i.e., $\Omega=2\sqrt{(\Delta_{\rm ge}-\Delta_{\rm es})\Delta_{\rm ge}}$, 
the scattering matrix becomes
\begin{eqnarray}
{\mathcal{S}_{\rm rot}}
=
\left[
  \begin{array}{cc}
    -\cos2\theta & -\sin2\theta \\
    -\sin2\theta & \cos2\theta \\
  \end{array}
\right].
\label{eq:S3}
\end{eqnarray}
A linearly polarized incident photon with polarization angle $\zeta$ ($0 \leq \zeta < \pi$), i.e., $|\Psi_{\rm in}\rangle=[\cos\zeta,\  \sin\zeta]^{\rm T}$ is scattered by the emitter and the output photon state is given by $|\Psi_{\rm out}\rangle=[\cos(2\theta-\zeta+\pi),\  \sin(2\theta-\zeta+\pi)]^{\rm T}$ which is also a linearly polarized photon whose polarization angle defined in the interval $[0,\pi]$ is given by
\begin{eqnarray}
\eta = \left\{
           \begin{array}{ll}
             2\theta-\zeta+\pi &  (-\pi \leq 2\theta-\zeta <0), \\
             2\theta-\zeta &  (0\leq 2\theta-\zeta <\pi), \\
             2\theta-\zeta-\pi & (\pi \leq 2\theta-\zeta < 2\pi). \\
           \end{array}
           \right.
           \label{eq:eta}
\end{eqnarray}
We can clearly see that by tuning the emitter dipole direction $\theta$, arbitrary rotation of a linearly polarized photon can be realized. Examples for the rotation of a horizontally linearly polarized photon are shown in Fig. S2 in the appendix \cite{sm}.

\begin{figure}[htb]
\centering
\includegraphics[width=0.48\columnwidth]{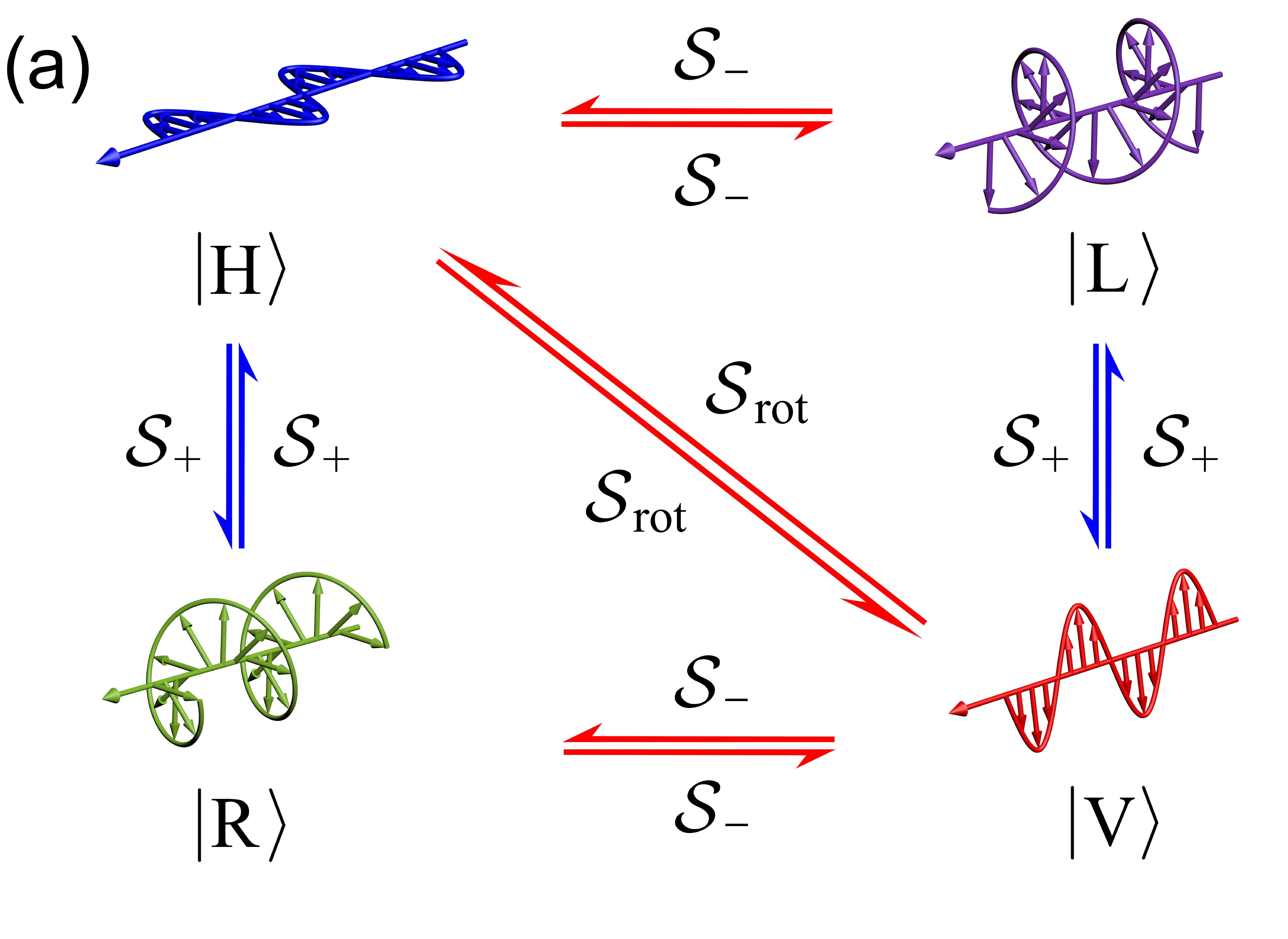}
\includegraphics[width=0.48\columnwidth]{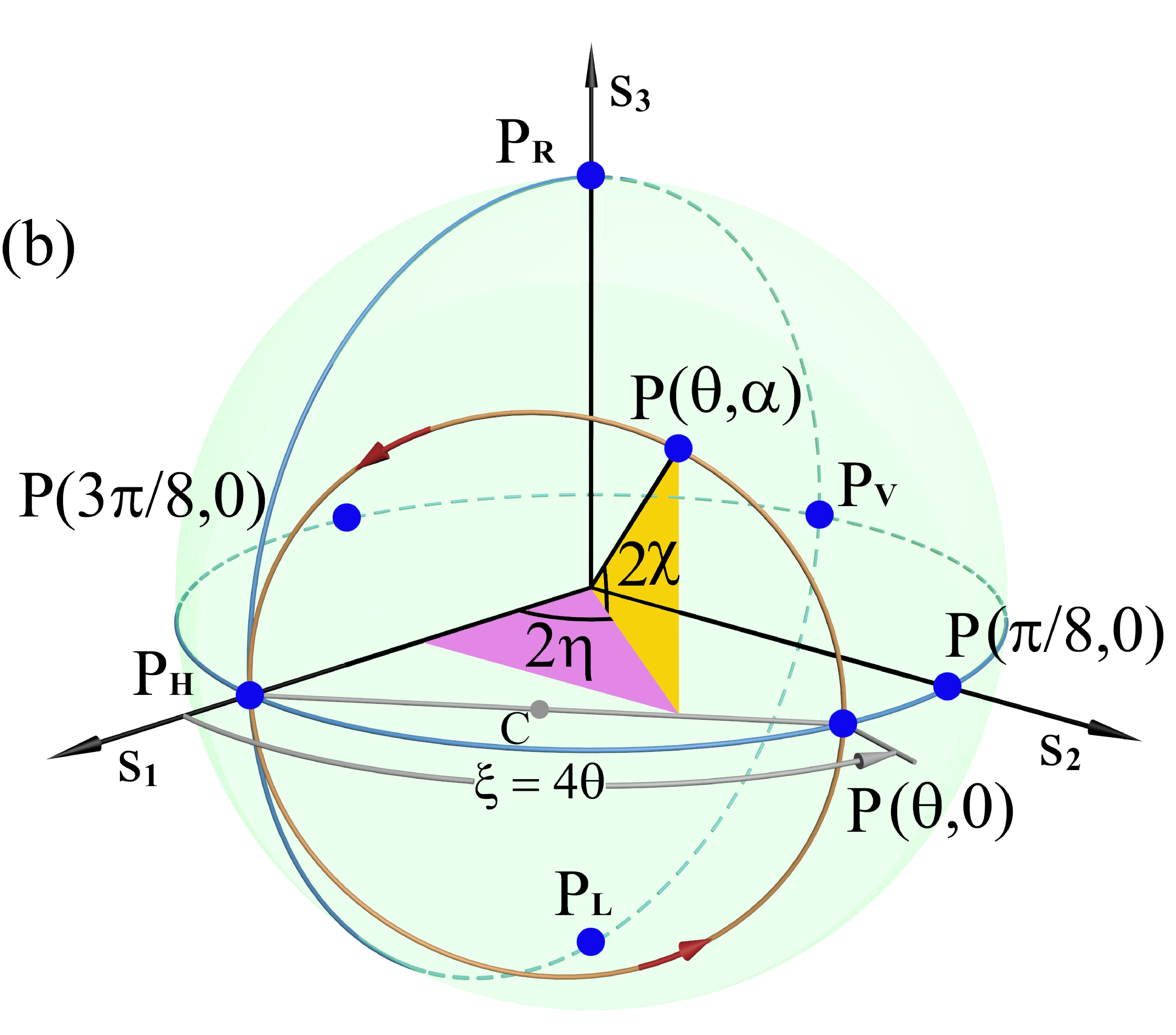}
\includegraphics[width=0.48\columnwidth]{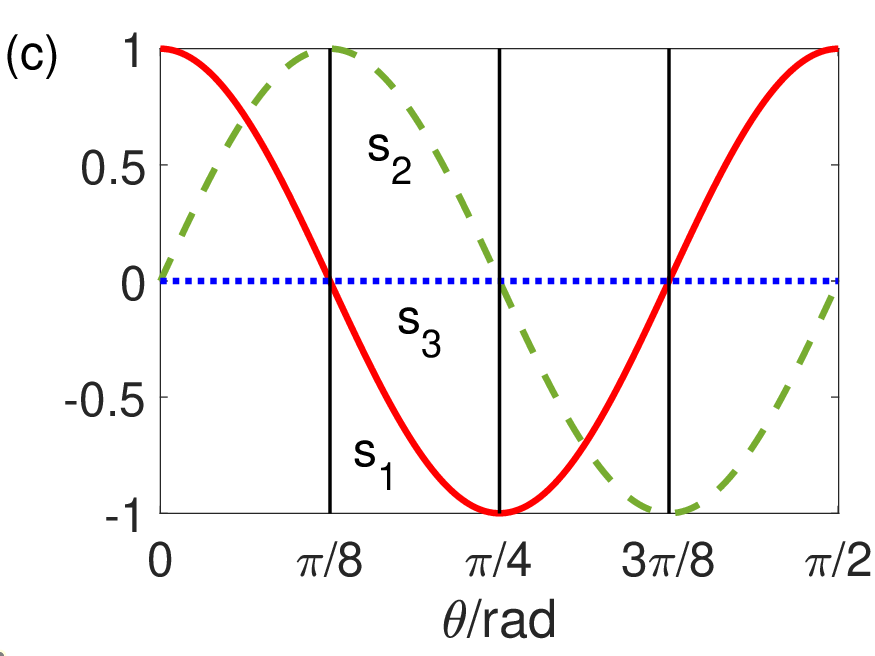}
\includegraphics[width=0.48\columnwidth]{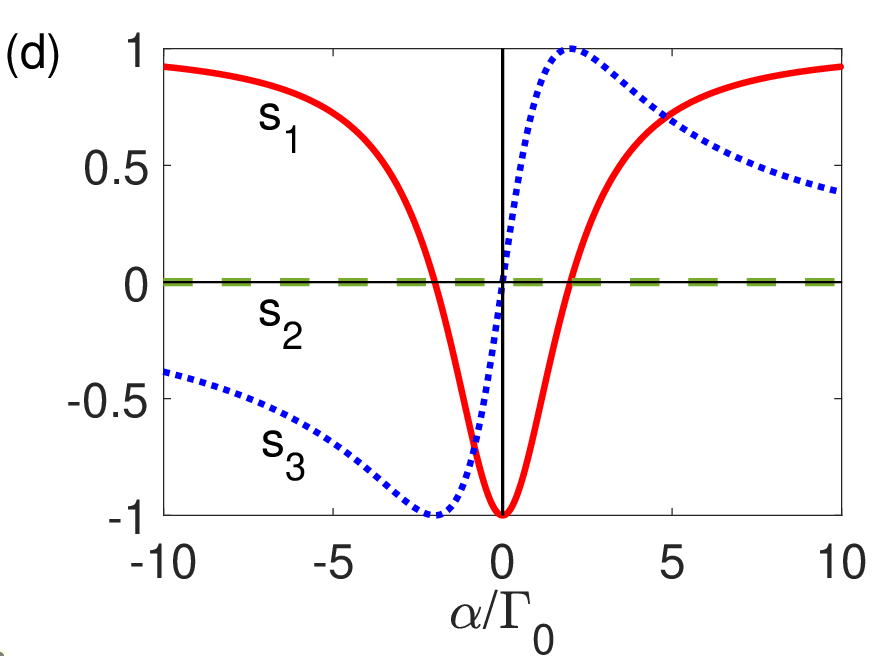}
\caption{(a) Conversion between a horizontally ($|\rm H\rangle$), vertically ($|\rm V\rangle$) polarized photon, and a left- ($|\rm L\rangle$), right- ($|\rm R\rangle$) handed circularly polarized photon through scatter matrices ${\mathcal{S}_+}$ and ${\mathcal{S}_-}$. (b) Poincar{\'e} sphere. 
(c) Stokes parameters $s_1$ (red solid curve), $s_2$ (green dashed curve) and $s_3$ (blue dotted curve) as functions of $\theta$ when $\alpha=0$.
(d) Stokes parameters $s_1$ (red solid curve), $s_2$ (green dashed curve) and $s_3$(blue dotted curve) as functions of $\alpha$ when $\theta=\pi/4$.
\label{fig:Poincare}}
\end{figure}

\textit{Arbitrary polarization in the Poincar{\'e} sphere-}Consider a photon with horizontal polarization is input into the system. 
By adjusting  $\theta$ and $\alpha$, we can obtain the output photon with arbitrary polarization states.
As shown in Fig.~\ref{fig:Poincare}(b), the polarization state of an output photon $|\Psi_{\rm out}\rangle=r_{\rm AA}(k)|\Psi^{(\rm A)}_{\rm R}\rangle+r_{\rm BA}(k)|\Psi^{(\rm B)}_{\rm R}\rangle$ can be represented by a point ${\rm P}(\theta,\alpha)$ on the Poincar{\'e} sphere~\cite{Born1999}.
Stokes parameters $\vec{s}$=($s_1$, $s_2$, $s_3$) are defined as $s_1=|r_{\rm AA}|^2-|r_{\rm BA}|^2$, $s_2=2|r_{\rm BA}||r_{\rm AA}| \cos\varphi$ and $s_3=2|r_{\rm BA}||r_{\rm AA}| \sin\varphi$, where $\varphi$ is the phase difference between $r_{\rm AA}$ and $r_{\rm BA}$. It is readily seen that $s_1^2+s_2^2+s_3^2=1$ if $|r_{\rm AA}(k)|^2+|r_{\rm BA}(k)|^2=1$. From Eqs. (1) and (2) without dissipation, it is straightforward to obtain
\begin{eqnarray}
s_1 &=& \frac{\cos4\theta+\alpha^2/(4\Gamma_0^2)}{1+\alpha^2/(4\Gamma_0^2)},\\
s_2 &=& \frac{\sin4\theta}{1+\alpha^2/(4\Gamma_0^2)},\\
s_3 &=& \frac{(\alpha/\Gamma_0)\sin2\theta}{1+\alpha^2/(4\Gamma_0^2)}.
\end{eqnarray}
In the Poincar{\'e} sphere, the angle $\eta=\tan^{-1}(s_2/s_1)/2$ ($0\leq\eta<\pi$) is equal to the included angle between the major axis of the polarization ellipse and the $x$-axis in the cross section of the waveguide. The angle $\chi=\tan^{-1}(s_3/\sqrt{s_1^2+s_2^2})/2$ ($-\pi/4\leq\chi\leq\pi/4$), and $\tan|\chi|$ is equal to the ratio of the minor axis to the major axis of the polarization ellipse. 
Now let us see how to set parameters $\theta$ and $\alpha$ to obtain an output photon with arbitrary polarization.
First, when $\alpha=\pm\infty$, i.e., $\Delta_{\rm ge}-\Delta_{\rm es}=0$ or $|\Delta_{\rm ge}-\Delta_{\rm es}|\ll\Omega^2$, $s_1=1$ and $s_2=s_3=0$ which is still horizontally polarized (${\rm P_{H}}$) in the Poincar{\'e} sphere due to the effect of EIT.
Second, when $\alpha=0$, $s_1=\cos4\theta, s_2=\sin4\theta, s_3=0$ and the output photon is represented by points on the equator ${\rm P(\theta,0)}$ which is linearly polarized with polarization angle $\eta=2\theta$.  With $\theta$ changes from $0$ to $\pi/2$, the Stokes parameters are shown in Fig.~\ref{fig:Poincare}(b) where points ${\rm P}(0,0)$, ${\rm P}(\pi/8,0)$, ${\rm P}(\pi/4,0)$ and ${\rm P}(3\pi/8,0)$ represent linear polarization angle $\eta=0$, $\pi/4$, $\pi/2$ and $3\pi/4$, respectively.
Third, for a given $\theta$, when $\alpha$ changes from $-\infty$ to $0$, and then to $+\infty$, the polarization ${\rm P}(\theta,\alpha)$ moves along the yellow circle, from the point ${\rm P_H}$, along the direction of the red arrows, through the point ${\rm P}(\theta,0)$, and finally back to the point ${\rm P_H}$. 
The plane of the circle is perpendicular to the $s_1$-$s_2$ plane,
and the circle has center $\vec{s}_{\rm c}=[\cos (4\theta)/2+1/2,\ \sin (4\theta)/2,\ 0]$ and radius $\sin 2\theta$.
Finally, when we adjust $\theta$ ($0\leq\theta<\pi/2$), the intersection point ${\rm P(\theta,0)}$ walks along the whole equator, 
and the yellow circle scans the whole Poincar{\'e} sphere. As an example, when $\theta=\pi/4$, point ${\rm P}(\theta,\alpha)$ is on the circle in the $s_1$-$s_3$ plane and when $\alpha$ changes from $-\infty$ to $+\infty$, the Stokes parameters are shown in Fig.~\ref{fig:Poincare}(c). Points ${\rm P}(\pi/4,-2\Gamma_0)$ and ${\rm P}(\pi/4,2\Gamma_0)$ represent the left- and right-handed circularly polarization, respectively.
From the above analysis, we can clearly see that by adjusting the parameters $\theta$ ($0\leq\theta<\pi/2$) and $\alpha$ ($-\infty<\alpha<+\infty$), the output polarization ${\rm P}(\theta,\alpha)$ can reach arbitrary point on the Poincar{\'e} sphere, which indicates that an output photon with arbitrary polarization can be generated.


\textit{Tunable working frequency-}The working frequency of our polarization converter is tunable. For an input photon whose frequency may significantly deviate from  the atomic resonant frequency, we can always tune the frequency $\nu$ and magnitude of the external control field to satisfy the conversion condition. 
As discussed in the previous sections, a specfic polarization conversion is described by a scattering matrix $\mathcal{S}$ which is a function of $\alpha$. Since $\alpha$ depends on $\Delta_{\rm ge}$, the scattering matrix $\mathcal{S}$ may change if the incident photon frequency varies. However, we can always tune the external driving field (i.e., $\Omega$ and $\Delta_{\rm es}$) to make $\alpha=\Omega^2/[4(\Delta_{\rm ge}-\Delta_{\rm es})]-\Delta_{\rm ge}$ unchanged so that the required conversion is faithfully realized (for details please see \cite{sm}). Therefore, our working frequency is broadband tunable which can find important applications in quantum photonic chips.

\begin{figure}[!ht]
\centering
\includegraphics[width=0.32\columnwidth]{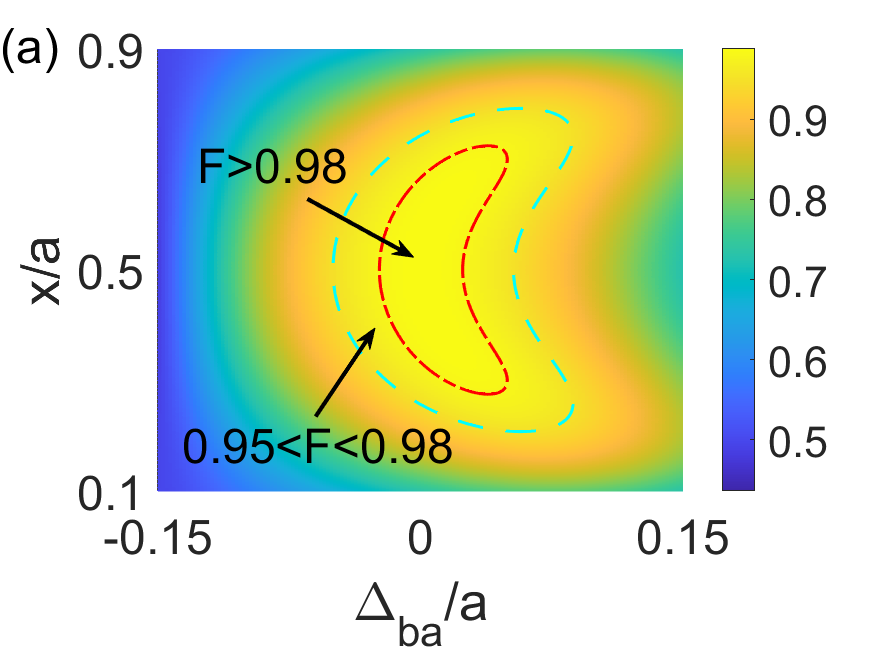}
\includegraphics[width=0.32\columnwidth]{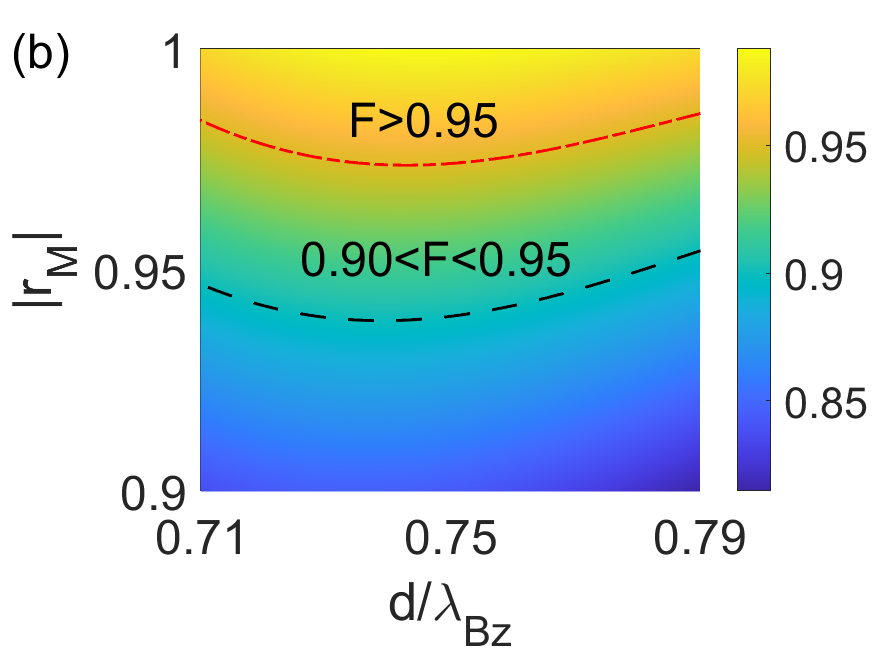}
\includegraphics[width=0.32\columnwidth]{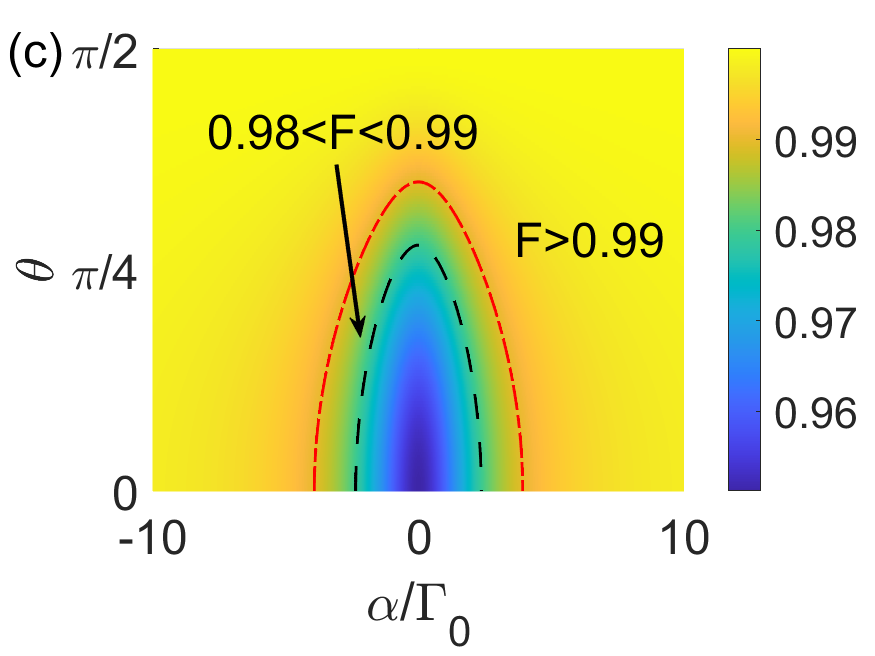}
\caption{(a,b) Fidelity of the polarization conversion from $|{\rm V}\rangle \to |{\rm L}\rangle$: (a) as functions of $\Delta_{ba}$ and the emitter lateral position $x$ variation with $y=b/2$, $d=0.75\lambda_{{\rm B}z}$, and $r_{\rm M}=-1$; (b) as functions of $d$ and reflectivity $r_{\rm M}$  with $x=a/2$, $y=b/2$, and $\Delta_{ba}=0$. (c) Conversion fidelity of a horizontal linearly polarized input light under different values of $\alpha$ and $\theta$ with external dissipation but other parameters are chosen as ideal values. For all three subfigures, external dissipation rate is $\gamma_{\rm e}=0.05\Gamma_{0}$. 
Here $\lambda_{{\rm B}z}=2\pi/k_{{\rm B}z}$, and $k_{{\rm B}z}=\sqrt{k^2-(\pi/a)^2}$ is the $z$-direction wave vector of mode B and $k=1.3\pi/a$.
\label{fig:nonideal}}
\end{figure}

\textit{Practical applicability-}Our scheme here is quite general. The requirements are that the waveguide has two degenerate modes with orthogonal polarization and the external dissipation should be much less than $\Gamma_0$. These requirements can be satisfied by a rectangular superconducting waveguide coupled to superconducting qubit \cite{Mirhosseini2019,Kumar2023} or a dielectric square waveguide such as photonic crystal waveguide coupled to cold atom \cite{Leong2020} or quantum dot \cite{Arcari2014,Scarpelli2019,Uppu2020}. In these systems, the $\beta-$factor (i.e., $\Gamma_0/(\gamma_{e}+\Gamma_{0})$) can be larger than $98\%$. 

In practice, we also need to consider the influence of nonideal conditions including the slight difference $\Delta_{ba}=b-a$ between the waveguide width $a$ and height $b$, the lateral (e.g. $x$) and longitudinal ($d$) position variation of the emitter, imperfect reflection coefficient $r_{\rm M}$ of the waveguide end, and external dissipation.
Without loss of generality, we consider the conversion of $|{\rm V}\rangle \to |{\rm L}\rangle$  under nonideal conditions and the result are shown in Fig.~\ref{fig:nonideal}.
In Fig.~\ref{fig:nonideal} (a), we present the conversion fidelity $F$ (the square of the modulus of the inner product between the result state and the target state) as a function of $\Delta_{ab}$ and $x$ with other parameters chosen as ideal values, from which we can see that the conversion efficiency can be larger than $95\%$ if $-0.05a\leq \Delta_{\rm ba}\leq 0.05a$ with $x=0.5a$ and can be larger than $98\%$ if $-0.022a\leq \Delta_{\rm ba}\leq 0.022a$ with $x=0.5a$ and $0.34a \leq x \leq 0.66a$ with $\Delta_{\rm ba}=0$.  For fixed value of $x$, the fidelity reduces as $|\Delta_{\rm ba}|$ increases because non-zero $|\Delta_{\rm ba}|$ leads to different wavelengths of mode A and mode B. Considering the mode A and mode B are standing waves in the semi-infinite waveguide, the two modes have different light intensities at the emitter's position, and their coupling strengths with the emitter deviate from the ideal values which leads to the reduction of the fidelity. When the emitter deviates from the center, the coupling strengths with two orthogonal modes deviate from the ideal values which leads to the reduction of the fidelity, but the result show that the conversion fidelity is still very high even if $x$ significantly deviates from the center ($F>98\%$ even if $0.34a<x<0.66a$ with $\Delta_{\rm ba}=0$). In Fig.~\ref{fig:nonideal} (b), we show the conversion fidelity as functions of $d$ and $r_{\rm M}$. It is clearly seen that the conversion fidelity does not decrease much when $d$ varies by certain small values.  When $0.71\lambda_{{\rm B}z}<d<0.79\lambda_{{\rm B}z}$ the conversion fidelity can still be larger than $95\%$ if $|r_{\rm M}|>0.986$. If the reflectivity of the mirror reduces, the fidelity decreases more obviously because the decrease of the reflection leads to the loss of the photon and the standing wave condition is violated. However, if the reflectivity is larger than $0.95$, the conversion fidelity is still larger than $90\%$. Finally, we also consider the effect of the external dissipation $\gamma_{e}$ and the results show that if $\gamma_{e}=0.05\Gamma_{0}$ which is achievable under current technology,  the conversion efficiency can still be larger than $95\%$. Especially when $\alpha$ is large, the conversion efficiency can still be larger than $99\%$ even if there is external dissipation due to electromagnetic induced transparency (EIT)-like effects.  Thus, our scheme still works well under certain nonideal conditions (for more general cases and more detail discussions, please see Sec. VII in \cite{sm}).

\textit{Conclusion-}We have proposed a chip-integrable scheme to realize arbitrary polarization conversion of single photons using a single atom coupled to a waveguide. 
In our scheme we can transform an input photon with arbitrary polarization into an output photon with any other polarization. The conversion efficiency can be unit in the ideal case and can still be larger than 90\% even if imperfect conditions are considered. In addition, the working frequency of our system can be adjusted continuously by tuning the strength and frequency of the external control field. Thus, our scheme here allows to manipulate the polarization degree of freedom conveniently on chip with high efficiency which allows to encode the polarization as qubits in the IQPC. Using polarziations instead of photonic paths as qubits may greatly reduce the size of the quantum photonic circuit and can find important applications in the integrated quantum device.

\begin{acknowledgments}
This work was supported by the National Key R\&D Program of China (Grant No. 2021YFA1400800), the Key Program of National Natural Science Foundation of China (Grant No. 12334017), the Key-Area Research and Development Program of Guangdong Province (Grant No.2018B030329001), the Guangdong Basic and Applied Basic Research Foundation (Grant No. 2023B1515040023), the Natural Science Foundations of Guangdong (Grant No.2021A1515010039), and Key Natural Scientific Research Projects of Universities in Anhui Province (2023AH051125).
\end{acknowledgments}

\appendix

\begin{widetext}

\section{$\rm TE_{01}$ and $\rm TE_{10}$ modes of a rectangular waveguide and the coupling with the emitter}

We consider ${\rm TE}_{01}$ mode and ${\rm TE}_{10}$ mode in a rectangular waveguide with cross section size $a\times b$
\begin{eqnarray}
E_x^{{(\rm A)}} &=& E_0 \sin\frac{{\pi y}}{b}{e^{i{k_z}z}},\\
H_y^{{(\rm A)}} &=& \frac{k_z}{\omega\mu} E_0 
\sin\frac{{\pi y}}{b}{e^{i{k_z}z}},\\
H_z^{{(\rm A)}} &=& \frac{i k_y}{\omega\mu} E_0 
\cos\frac{{\pi y}}{b}{e^{i{k_z}z}},\\
E_y^{{(\rm B)}} &=& E_0 
\sin\frac{{\pi x}}{a}{e^{i{k_z}z}},\\
H_x^{{(\rm B)}} &=& -\frac{k_z}{\omega\mu} E_0 
\sin\frac{{\pi x}}{a}{e^{i{k_z}z}},\\
H_z^{{(\rm B)}} &=& -\frac{i k_x}{\omega\mu} E_0 
\cos\frac{{\pi x}}{a}{e^{i{k_z}z}},
\end{eqnarray}
and $E_y^{(\rm A)}=E_z^{{(\rm A)}}=H_x^{{(\rm A)}}=E_x^{{(\rm B)}}=E_z^{{(\rm B)}}=H_y^{{(\rm B)}}=0$. 
Here superscripts ``(A)" and ``(B)" denote ${\rm TE_{01}}$ mode and ${\rm TE_{10}}$ mode, respectively.
$E_0=\sqrt{\hbar\omega/(2\epsilon V_{\rm eff})}$ and $V_{\rm eff}$ is the mode volume.
$\epsilon$ and $\mu$ are the dielectric constant, permeability of the waveguide, respectively.
$\omega$ and $k_z$ are the angular frequency and the z component of the wave vector of the electromagnetic wave, respectively.


\begin{figure}[!ht]
\centering
\includegraphics[width=0.3\columnwidth]{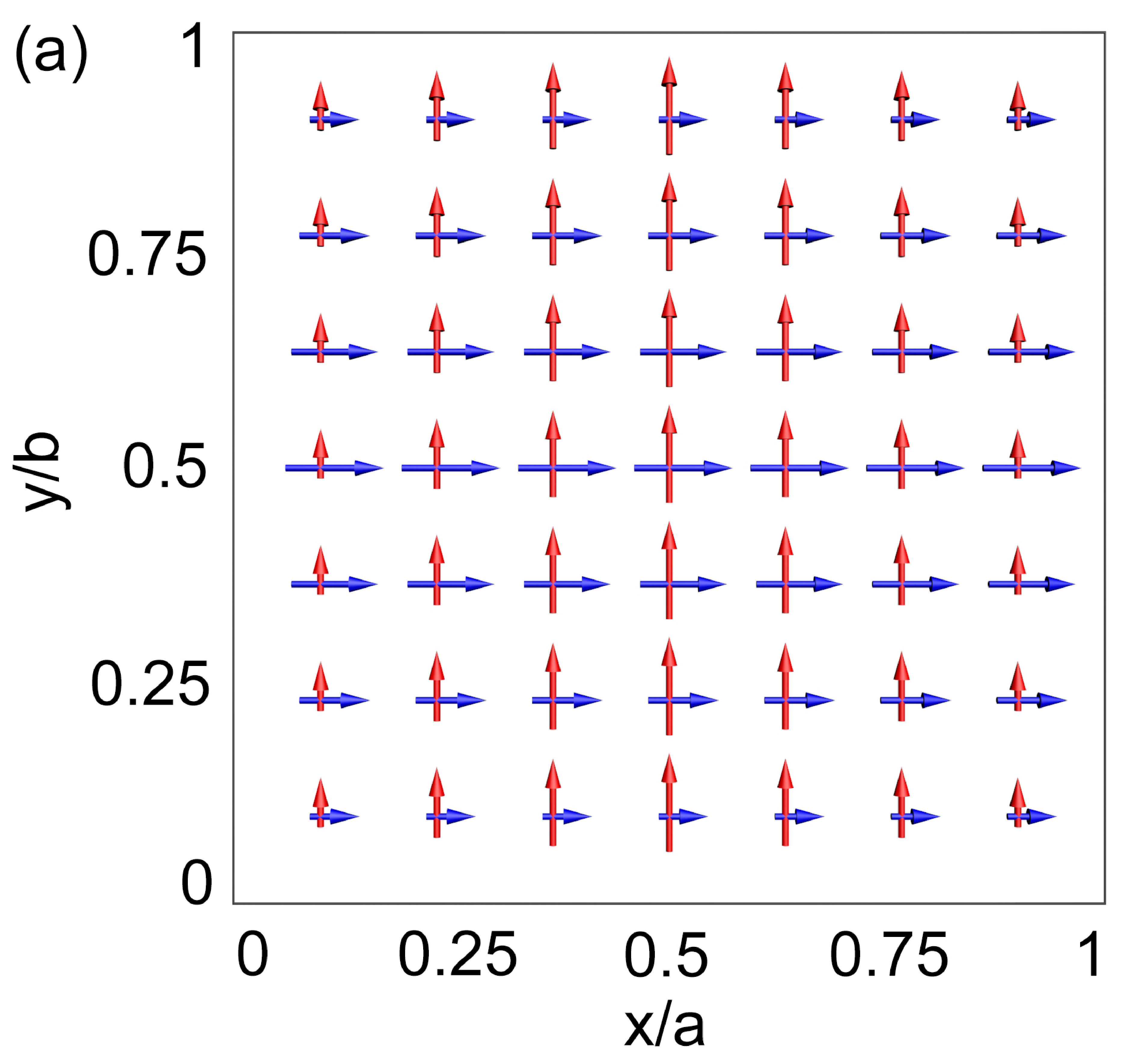}
\includegraphics[width=0.3\columnwidth]{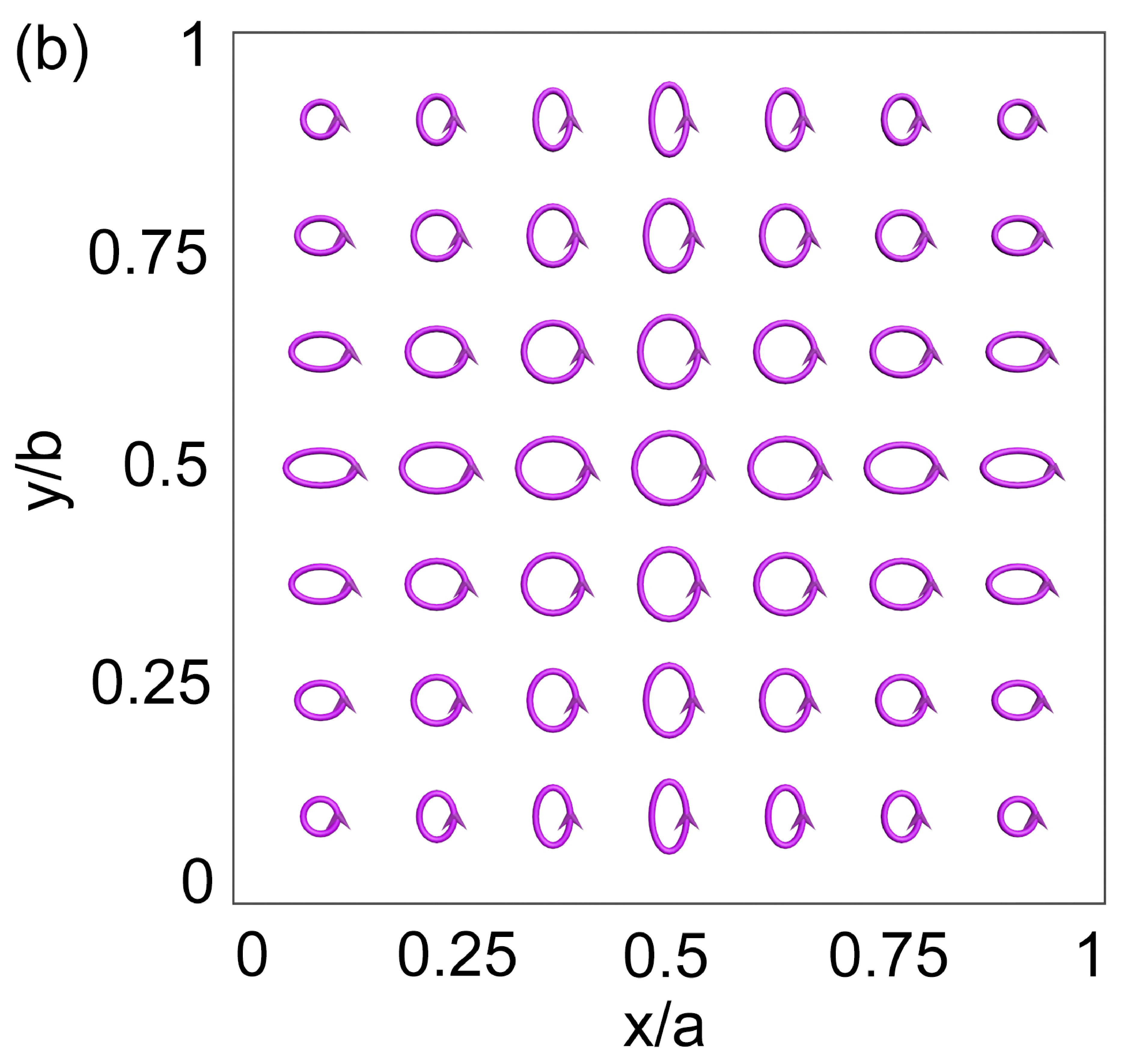}
\caption{(a) Electric fields $\vec{E}_{01}$ of mode ${\rm TE_{01}}$ (blue arrows) and electric fields $\vec{E}_{10}$ of mode ${\rm TE_{10}}$ (red arrows) in the cross section of waveguide.
(b) The superposition of modes ${\rm TE_{01}}$ and ${\rm TE_{10}}$ with phase difference $\pi/2$ leads to left-handed circularly polarized field at the center and along the two diagonals of the cross section of the waveguide, and elliptically polarized field at other position. 
\label{fig:Polarization}}
\end{figure}

In Fig.~\ref{fig:Polarization}(a), the electric fields of the horizontally polarized light and the vertically polarized light are represented with blue and red arrows, respectively.
Fig.~\ref{fig:Polarization}(b) shows the electric fields of the left-handed elliptically polarized light. At the center and along the diagonals of the cross section of the waveguide, the light is circularly polarized.

The coupling strength between the emitter and the electromagnetic field of photon A mode in three-dimensional space is~\cite{Scully1997}
\begin{eqnarray}
g_{\rm A}=-\frac{\mu_{\rm ge} |E_x^{(\rm A)}| {\rm cos}\theta}{\hbar}=-\mu_{\rm ge}\sqrt{\frac{\omega}{2\hbar\epsilon V_{\rm eff}}} 
\sin\frac{\pi y}{b} \cos\theta.
\end{eqnarray}
Then we consider the waveguide as a one-dimensional continual space, the coupling strength and the emission rate of the emitter are
\begin{eqnarray}
V_{\rm A}
&=&\sqrt{L}g_{\rm A}
=-\mu_{\rm ge}\sqrt{\frac{\omega}{2\hbar \epsilon S_{\rm eff}}} \sin\frac{{\pi y}}{b} \cos\theta,\\
\Gamma_{\rm A}
&=&\frac{2V_{\rm A}^2}{v_{\rm g}}
=\frac{\mu_{\rm ge}^2\omega}{\hbar\epsilon v_{\rm g} S_{\rm eff}} \sin^2\frac{{\pi y}}{b} {\cos}^2\theta.
\end{eqnarray}
Similarly, the coupling strength between the emitter and photon modes B and the emission rate of the emitter are
\begin{eqnarray}
g_{\rm B}&=&-\frac{\mu_{\rm ge} |E_y^{(\rm B)}| {\rm sin}\theta}{\hbar}=-\mu_{\rm ge}\sqrt{\frac{\omega}{2\hbar\epsilon V_{\rm eff}}} 
\sin\frac{\pi x}{a}\sin \theta,\\
V_{\rm B}
&=& \sqrt{L}g_{\rm B}
=-\mu_{\rm ge}\sqrt{\frac{\omega}{2\hbar \epsilon S_{\rm eff}}} \sin\frac{{\pi x}}{a} \sin\theta,\\
\Gamma_{\rm B}&=&\frac{2V_{\rm B}^2}{v_{\rm g}}
=\frac{\mu_{\rm ge}^2\omega}{\hbar\epsilon v_{\rm g} S_{\rm eff}} \sin^2\frac{{\pi x}}{a} {\sin}^2\theta.
\end{eqnarray}
Here, $\mu_{\rm ge}$ is the electric dipole moment of the transition $|{\rm g}\rangle \to |{\rm e}\rangle$ of the emitter. $V_{\rm eff}$, $L$ and $S_{\rm eff}$ are the effective mode volume, the effective length and the effective cross sectional area of the waveguide, respectively.

We define
\begin{eqnarray}
\Gamma_{0}\equiv \frac{\mu_{\rm ge}^2\omega}{2\hbar\epsilon v_{\rm g} S_{\rm eff}}
\end{eqnarray}
as a unit, and $\Gamma_{\rm A}$ and $\Gamma_{\rm B}$ can be written as
\begin{subequations}
\begin{eqnarray}
\Gamma_{\rm A} &=& 2\Gamma_{0}\sin^2\frac{{\pi y}}{b} \cos^2\theta,\\
\Gamma_{\rm B} &=& 2\Gamma_{0}\sin^2\frac{{\pi x}}{a} \sin^2\theta.
\end{eqnarray}
\label{eq:Gamma_a_b}
\end{subequations}
In this way, when the emitter is at the point $(x,y)=(a/2,b/2)$ in the cross section, and its electric dipole moment direction is $\theta=\pi/4$, we have 
$\Gamma_{\rm A}=\Gamma_{\rm B}=\Gamma_0$.

\section{Derivation of the scattering matrix elements $r_{\rm AA}$, $r_{\rm BA}$, $r_{\rm AB}$, and $r_{\rm BB}$}

We assume the input single photon is a monochromatic wave in mode A
\begin{eqnarray}
|\Psi^{(\rm A)}_{\rm L}\rangle = \int{\rm d}z \frac{1}{\sqrt{2\pi}}e^{-ik_{\rm A}z} a_{\rm L}^{\dagger}(z) |{\rm g},0\rangle.
\end{eqnarray}
Here, $k_{\rm A}$ is the $z$-direction wave vector of mode A.
The scattering eigenstate can be written in the form~\cite{JTShen2009,Roy2017,HuaixiuZheng2013,Bradford2012PRL,Burillo2016}
\begin{eqnarray}
|\Psi^{(\rm A)}_{\rm eig}\rangle 
=&& \int{\rm d}z f^{\rm (A)}_{\rm AL}(k_{\rm A},z) a_{\rm L}^{\dagger}(z) |{\rm g},0\rangle
+\int{\rm d}z f^{\rm (A)}_{\rm AR}(k_{\rm A},z) a_{\rm R}^{\dagger}(z) |{\rm g},0\rangle
+ \int{\rm d}z f^{\rm (A)}_{\rm BL}(k_{\rm B},z) b_{\rm L}^{\dagger}(z) |{\rm g},0\rangle\nonumber\\
&&+\int{\rm d}z f^{\rm (A)}_{\rm BR}(k_{\rm B},z) b_{\rm R}^{\dagger}(z) |{\rm g},0\rangle
+ c^{\rm (A)}_{\rm e}(\omega) |{\rm e},0\rangle
+ c^{\rm (A)}_{\rm s}(\omega) |{\rm s},0\rangle.
\label{eq:Psi_three}
\end{eqnarray}
Upon substituting Hamiltonian $H$ and state (\ref{eq:Psi_three}) into the Schr{\"o}dinger equation
\begin{eqnarray}
H|\Psi^{(\rm A)}_{\rm eig}\rangle = \omega|\Psi^{(\rm A)}_{\rm eig}\rangle,
\end{eqnarray}
we obtain the following equations
\begin{eqnarray}
(\omega_{\rm 0A}+iv_{\rm gA} \frac{\rm d}{{\rm d}z})f^{\rm (A)}_{\rm AL}(k_{\rm A},z)+V_{\rm A}\delta(z-z_0)c^{\rm (A)}_{\rm e}(\omega) &=& \omega f^{\rm (A)}_{\rm AL}(k_{\rm A},z),\label{eq:Sch_three_1}\\
(\omega_{\rm 0A}-iv_{\rm gA} \frac{\rm d}{{\rm d}z})f^{\rm (A)}_{\rm AR}(k_{\rm A},z)+V_{\rm A}\delta(z-z_0)c^{\rm (A)}_{\rm e}(\omega) &=& \omega f^{\rm (A)}_{\rm AR}(k_{\rm A},z),\label{eq:Sch_three_2}\\
(\omega_{\rm 0B}+iv_{\rm gB} \frac{\rm d}{{\rm d}z})f^{\rm (A)}_{\rm BL}(k_{\rm B},z)+V_{\rm B}\delta(z-z_0)c^{\rm (A)}_{\rm e}(\omega) &=& \omega f^{\rm (A)}_{\rm BL}(k_{\rm B},z),\label{eq:Sch_three_3}\\
(\omega_{\rm 0B}-iv_{\rm gB} \frac{\rm d}{{\rm d}z})f^{\rm (A)}_{\rm BR}(k_{\rm B},z)+V_{\rm B}\delta(z-z_0)c^{\rm (A)}_{\rm e}(\omega) &=& \omega f^{\rm (A)}_{\rm BR}(k_{\rm B},z),\label{eq:Sch_three_4}\\
V_{\rm A}f^{\rm (A)}_{\rm AL}(k_{\rm A},z_0) 
+ V_{\rm A}f^{\rm (A)}_{\rm AR}(k_{\rm A},z_0) 
+ V_{\rm B}f^{\rm (A)}_{\rm BL}(k_{\rm B},z_0)&+& V_{\rm B}f^{\rm (A)}_{\rm BR}(k_{\rm B},z_0)\nonumber\\
+(\omega_{\rm e}-i\gamma_{\rm e}/2)c^{\rm (A)}_{\rm e}(\omega) 
+ \frac{\Omega}{2}c^{\rm (A)}_{\rm s}(\omega)
&=& \omega c^{\rm (A)}_{\rm e}(\omega),\label{eq:Sch_three_5}\\
(\omega_{\rm e}-\Delta_{\rm es})c^{\rm (A)}_{\rm s}(\omega) + \frac{\Omega}{2}c^{\rm (A)}_{\rm e}(\omega) &=& \omega c^{\rm (A)}_{\rm s}(\omega).\label{eq:Sch_three_6}
\end{eqnarray}
We make the following ansatz on the amplitudes
\begin{eqnarray}
f^{\rm (A)}_{\rm AL}(k_{\rm A},z) &=& \frac{1}{\sqrt{2\pi}}e^{-ik_{\rm A}z}[t^{\rm (A)}_{\rm A}(\omega)\theta(-z+z_0)+\theta(z-z_0)],\label{eq:ansatz_three_1}\\
f^{\rm (A)}_{\rm AR}(k_{\rm A},z) &=& \frac{1}{\sqrt{2\pi}}e^{ik_{\rm A}z}[r^{\rm (A)}_{\rm A}(\omega)\theta(-z+z_0)+r_{\rm AA}(\omega)\theta(z-z_0)],\label{eq:ansatz_three_2}\\
f^{\rm (A)}_{\rm BL}(k_{\rm B},z) &=& \frac{1}{\sqrt{2\pi}}e^{-ik_{\rm B}z}t^{\rm (A)}_{\rm B}(\omega)\theta(-z+z_0),\\
f^{\rm (A)}_{\rm BR}(k_{\rm B},z) &=& \frac{1}{\sqrt{2\pi}}e^{ik_{\rm B}z}[r^{\rm (A)}_{\rm B}(\omega)\theta(-z+z_0)+r_{\rm BA}(\omega)\theta(z-z_0)].\label{eq:ansatz_three_4}
\end{eqnarray}
Inserting the ansatz (\ref{eq:ansatz_three_1})-(\ref{eq:ansatz_three_4}) into Eqs. (\ref{eq:Sch_three_1})-(\ref{eq:Sch_three_6}), we obtain equations
\begin{eqnarray}
iv_{\rm gA}\frac{1}{\sqrt{2\pi}} e^{-ik_{\rm A}z_0} [1-t^{\rm (A)}_{\rm A}(\omega)] + V_{\rm A}c^{\rm (A)}_{\rm e}(\omega) &=& 0,\label{eq:Sch_three_7}\\
-iv_{\rm gA}\frac{1}{\sqrt{2\pi}} e^{ik_{\rm A}z_0} [r_{\rm AA}(\omega)-r^{\rm (A)}_{\rm A}(\omega)] + V_{\rm A}c^{\rm (A)}_{\rm e}(\omega) &=& 0,\label{eq:Sch_three_8}\\
iv_{\rm gB}\frac{1}{\sqrt{2\pi}} e^{-ik_{\rm B}z_0} [-t^{\rm (A)}_{\rm B}(\omega)] + V_{\rm B}c^{\rm (A)}_{\rm e}(\omega) &=& 0,\label{eq:Sch_three_9}\\
-iv_{\rm gB}\frac{1}{\sqrt{2\pi}} e^{ik_{\rm B}z_0} [r_{\rm BA}(\omega)-r^{\rm (A)}_{\rm B}(\omega)] + V_{\rm B}c^{\rm (A)}_{\rm e}(\omega) &=& 0,\label{eq:Sch_three_10}\\
\frac{e^{-ik_{\rm A}z_0}}{\sqrt{2\pi}} \frac{1+t^{\rm (A)}_{\rm A}(\omega)}{2}V_{\rm A}
+\frac{e^{ik_{\rm A}z_0}}{\sqrt{2\pi}} \frac{r_{\rm AA}(\omega)+r^{\rm (A)}_{\rm A}(\omega)}{2}V_{\rm A} 
+\frac{e^{-ik_{\rm B}z_0}}{\sqrt{2\pi}} \frac{t^{\rm (A)}_{\rm B}(\omega)}{2}V_{\rm B} &&\nonumber\\
+\frac{e^{ik_{\rm B}z_0}}{\sqrt{2\pi}} \frac{r_{\rm BA}(\omega)+r^{\rm (A)}_{\rm B}(\omega)}{2}V_{\rm B} 
+\frac{\Omega}{2}c^{\rm (A)}_{\rm s}(\omega)
- (\omega-\omega_{\rm e}+i\gamma_{\rm e}/2)c^{\rm (A)}_{\rm e}(\omega)&=&0,\label{eq:Sch_three_11}\\
\frac{\Omega}{2}c^{\rm (A)}_{\rm e}(\omega) - (\omega-\omega_{\rm e}+\Delta_{\rm es})c^{\rm (A)}_{\rm s}(\omega)&=&0.
\label{eq:Sch_three_12}
\end{eqnarray}
The boundary condition at the end of the waveguide is
\begin{eqnarray}
f^{\rm (A)}_{\rm AR}(k_{\rm A},z_{\rm M})&=&r_{\rm AM} f^{\rm (A)}_{\rm AL}(k_{\rm A},z_{\rm M}),\label{eq:Sch_three_13}\\
f^{\rm (A)}_{\rm BR}(k_{\rm B},z_{\rm M})&=&r_{\rm BM} f^{\rm (A)}_{\rm BL}(k_{\rm B},z_{\rm M}),\label{eq:Sch_three_14}
\end{eqnarray}
where $z_{\rm M}$ is the $z$ coordinate of the mirror at the end of the waveguide.
Solving Eqs.~(\ref{eq:Sch_three_7})-(\ref{eq:Sch_three_14}), we obtain the coefficients
\begin{eqnarray}
t^{\rm (A)}_{\rm A}(\omega) &=& \frac{\alpha+i\gamma_{\rm e}/2+i(1+r_{\rm BM} e^{i\phi_{\rm B}})\Gamma_{\rm B}/2}
{\alpha+i\gamma_{\rm e}/2+i(1+r_{\rm AM} e^{i\phi_{\rm A}})\Gamma_{\rm A}/2+i(1+r_{\rm BM} e^{i\phi_{\rm B}})\Gamma_{\rm B}/2},\\
r^{\rm (A)}_{\rm A}(\omega) &=&  \frac{\alpha+i\gamma_{\rm e}/2+i(1+r_{\rm BM} e^{i\phi_{\rm B}})\Gamma_{\rm B}/2}
{\alpha+i\gamma_{\rm e}/2+i(1+r_{\rm AM} e^{i\phi_{\rm A}})\Gamma_{\rm A}/2+i(1+r_{\rm BM} e^{i\phi_{\rm B}})\Gamma_{\rm B}/2}
r_{\rm AM}e^{-2ik_{\rm A}z_{\rm M}},\\
r_{\rm AA}(\omega) &=&  \frac{\alpha+i\gamma_{\rm e}/2-i(1+e^{-i\phi_{\rm A}}/r_{\rm AM})\Gamma_{\rm A}/2+i(1+e^{i\phi_{\rm B}}r_{\rm BM})\Gamma_{\rm B}/2}
{\alpha+i\gamma_{\rm e}/2+i(1+r_{\rm AM} e^{i\phi_{\rm A}})\Gamma_{\rm A}/2+i(1+r_{\rm BM} e^{i\phi_{\rm B}})\Gamma_{\rm B}/2}
r_{\rm AM}e^{-2ik_{\rm A}z_{\rm M}},\label{eq:tAA_0}\\
t^{\rm (A)}_{\rm B}(\omega) &=& \frac{-i(1+r_{\rm AM}e^{i\phi_{\rm A}}) (\Gamma_{\rm B}/2)(V_{\rm A}/V_{\rm B})}
{\alpha+i\gamma_{\rm e}/2+i(1+r_{\rm AM} e^{i\phi_{\rm A}})\Gamma_{\rm A}/2+i(1+r_{\rm BM} e^{i\phi_{\rm B}})\Gamma_{\rm B}/2} e^{i(k_{\rm B}-k_{\rm A})z_0},\\
r^{\rm (A)}_{\rm B}(\omega) &=&  \frac{-i(1+r_{\rm AM}e^{i\phi_{\rm A}}) (\Gamma_{\rm B}/2)(V_{\rm A}/V_{\rm B})}
{\alpha+i\gamma_{\rm e}/2+i(1+r_{\rm AM} e^{i\phi_{\rm A}})\Gamma_{\rm A}/2+i(1+r_{\rm BM} e^{i\phi_{\rm B}})\Gamma_{\rm B}/2} e^{i(k_{\rm B}-k_{\rm A})z_0}
r_{\rm BM}e^{-2ik_{\rm B}z_{\rm M}},\\
r_{\rm BA}(\omega) &=&  \frac{-i[1+r_{\rm AM}e^{i\phi_{\rm A}}+e^{-i\phi_{\rm B}}/r_{\rm BM}+e^{i(\phi_{\rm A}-\phi_{\rm B})}r_{\rm AM}/r_{\rm BM}] (\Gamma_{\rm B}/2)(V_{\rm A}/V_{\rm B})}
{\alpha+i\gamma_{\rm e}/2+i(1+r_{\rm AM} e^{i\phi_{\rm A}})\Gamma_{\rm A}/2+i(1+r_{\rm BM} e^{i\phi_{\rm B}})\Gamma_{\rm B}/2}
e^{i(k_{\rm B}-k_{\rm A})z_0}
r_{\rm BM}e^{-2ik_{\rm B}z_{\rm M}},\label{eq:tBA_0}\\
c^{\rm (A)}_{\rm e}(\omega) &=& \frac{e^{-ik_{\rm A}z_0}(1+r_{\rm AM}e^{i\phi_{\rm A}})V_{\rm A}/\sqrt{2\pi}}
{\alpha+i\gamma_{\rm e}/2+i(1+r_{\rm AM} e^{i\phi_{\rm A}})\Gamma_{\rm A}/2+i(1+r_{\rm BM} e^{i\phi_{\rm B}})\Gamma_{\rm B}/2},\\
c^{\rm (A)}_{\rm s}(\omega) &=& \frac{e^{-ik_{\rm A}z_0}(1+r_{\rm AM}e^{i\phi_{\rm A}})V_{\rm A}\Omega/[2\sqrt{2\pi}(\omega-\omega_{\rm e}+\Delta_{\rm es})]}
{\alpha+i\gamma_{\rm e}/2+i(1+r_{\rm AM} e^{i\phi_{\rm A}})\Gamma_{\rm A}/2+i(1+r_{\rm BM} e^{i\phi_{\rm B}})\Gamma_{\rm B}/2}.
\end{eqnarray}
Here, we define $\alpha=\Omega^2/[4(\Delta_{\rm ge}-\Delta_{\rm es})]-\Delta_{\rm ge}$, and 
$\Delta_{\rm ge}=\omega_{\rm e}-\omega$ is the detuning between the emitter transition $|{\rm g}\rangle \to |{\rm e}\rangle$ and the input photon. 
$\Gamma_{\rm A}=2V_{\rm A}^2/v_{\rm gA}$ and $\Gamma_{\rm B}=2V_{\rm B}^2/v_{\rm gB}$ are the rates that the emitter emits mode-A photon and mode-B photon through the transition $|{\rm e}\rangle\to|{\rm g}\rangle$, respectively.
We define phase $\phi_{\rm A}=2k_{\rm A}d$ and $\phi_{\rm B}=2k_{\rm B}d$, where $d=z_{\rm 0}-z_{\rm M}$ is the separation between the emitter and the end of the waveguide. According to the Lippmann-Schwinger formalism~\cite{Sakurai1994,JTShen2007}, we can obtain the output state
\begin{eqnarray}
|\Psi^{(\rm A)}_{\rm out}\rangle 
&=& r_{\rm AA}(\omega) \int{\rm d}z \frac{e^{ik_{\rm A}z}}{\sqrt{2\pi}} a_{\rm R}^{\dagger}(z) |{\rm g},0\rangle
+r_{\rm BA}(\omega) \int{\rm d}z \frac{e^{ik_{\rm B}z}}{\sqrt{2\pi}} b_{\rm R}^{\dagger}(z) |{\rm g},0\rangle\nonumber\\
&=& r_{\rm AA}(\omega) |\Psi^{(\rm A)}_{\rm R}\rangle + r_{\rm BA}(\omega) |\Psi^{(\rm B)}_{\rm R}\rangle.
\end{eqnarray}

When the input photon is in mode B,
\begin{eqnarray}
|\Psi^{(\rm B)}_{\rm L}\rangle = \int{\rm d}z \frac{1}{\sqrt{2\pi}}e^{-ik_{\rm B}z} b_{\rm L}^{\dagger}(z) |{\rm g},0\rangle,
\end{eqnarray}
the scattering eigenstate is
\begin{eqnarray}
|\Psi^{(\rm B)}_{\rm eig}\rangle 
=&& \int{\rm d}z f^{\rm (B)}_{\rm AL}(k_{\rm A},z) a_{\rm L}^{\dagger}(z) |{\rm g},0\rangle
+ \int{\rm d}z f^{\rm (B)}_{\rm AR}(k_{\rm A},z) a_{\rm R}^{\dagger}(z) |{\rm g},0\rangle
+ \int{\rm d}z f^{\rm (B)}_{\rm BL}(k_{\rm B},z) b_{\rm L}^{\dagger}(z) |{\rm g},0\rangle\nonumber\\
&&+\int{\rm d}z f^{\rm (B)}_{\rm BR}(k_{\rm B},z) b_{\rm R}^{\dagger}(z) |{\rm g},0\rangle
+ c^{\rm (B)}_{\rm e}(\omega) |{\rm e},0\rangle
+ c^{\rm (B)}_{\rm s}(\omega) |{\rm s},0\rangle,
\end{eqnarray}
where
\begin{eqnarray}
f^{\rm (B)}_{\rm AL}(k_{\rm A},z) &=& \frac{1}{\sqrt{2\pi}}e^{-ik_{\rm A}z}t^{\rm (B)}_{\rm A}(\omega)\theta(-z+z_0),\\
f^{\rm (B)}_{\rm AR}(k_{\rm A},z) &=& \frac{1}{\sqrt{2\pi}}e^{ik_{\rm A}z}[r^{\rm (B)}_{\rm A}(\omega)\theta(-z+z_0)+r_{\rm AB}(\omega)\theta(z-z_0)],\\
f^{\rm (B)}_{\rm BL}(k_{\rm B},z) &=& \frac{1}{\sqrt{2\pi}}e^{-ik_{\rm B}z}[t^{\rm (B)}_{\rm B}(\omega)\theta(-z+z_0)+\theta(z-z_0)],\\
f^{\rm (B)}_{\rm BR}(k_{\rm B},z) &=& \frac{1}{\sqrt{2\pi}}e^{ik_{\rm B}z}[r^{\rm (B)}_{\rm B}(\omega)\theta(-z+z_0)+r_{\rm BB}(\omega)\theta(z-z_0)].
\end{eqnarray}
and
\begin{eqnarray}
t^{\rm (B)}_{\rm A}(\omega) &=& \frac{-i(1+r_{\rm BM}e^{i\phi_{\rm B}})(\Gamma_{\rm A}/2)(V_{\rm B}/V_{\rm A})}
{\alpha+i\gamma_{\rm e}/2+i(1+r_{\rm AM} e^{i\phi_{\rm A}})\Gamma_{\rm A}/2+i(1+r_{\rm BM} e^{i\phi_{\rm B}})\Gamma_{\rm B}/2}
e^{i(k_{\rm A}-k_{\rm B})z_0},\\
r^{\rm (B)}_{\rm A}(\omega) &=&  \frac{-i(1+r_{\rm BM}e^{i\phi_{\rm B}})(\Gamma_{\rm A}/2)(V_{\rm B}/V_{\rm A})}
{\alpha+i\gamma_{\rm e}/2+i(1+r_{\rm AM} e^{i\phi_{\rm A}})\Gamma_{\rm A}/2+i(1+r_{\rm BM} e^{i\phi_{\rm B}})\Gamma_{\rm B}/2}
e^{i(k_{\rm A}-k_{\rm B})z_0}
r_{\rm AM}e^{-2ik_{\rm A}z_{\rm M}},\\
r_{\rm AB}(\omega) &=&  \frac{-i[1+e^{i\phi_{\rm B}}r_{\rm BM}+e^{-i\phi_{\rm A}}/r_{\rm AM}+e^{i(\phi_{\rm B}-\phi_{\rm A})}r_{\rm BM}/r_{\rm AM}](\Gamma_{\rm A}/2)(V_{\rm B}/V_{\rm A})}
{\alpha+i\gamma_{\rm e}/2+i(1+r_{\rm AM} e^{i\phi_{\rm A}})\Gamma_{\rm A}/2+i(1+r_{\rm BM} e^{i\phi_{\rm B}})\Gamma_{\rm B}/2}
e^{i(k_{\rm A}-k_{\rm B})z_0}
r_{\rm AM}e^{-2ik_{\rm A}z_{\rm M}},\label{eq:tAB_0}\\
t^{\rm (B)}_{\rm B}(\omega) &=& \frac{\alpha+i\gamma_{\rm e}/2+i(1+r_{\rm AM}e^{i\phi_{\rm A}})\Gamma_{\rm A}/2}
{\alpha+i\gamma_{\rm e}/2+i(1+r_{\rm AM} e^{i\phi_{\rm A}})\Gamma_{\rm A}/2+i(1+r_{\rm BM} e^{i\phi_{\rm B}})\Gamma_{\rm B}/2},\\
r^{\rm (B)}_{\rm B}(\omega) &=&  \frac{\alpha+i\gamma_{\rm e}/2+i(1+r_{\rm AM}e^{i\phi_{\rm A}})\Gamma_{\rm A}/2}
{\alpha+i\gamma_{\rm e}/2+i(1+r_{\rm AM} e^{i\phi_{\rm A}})\Gamma_{\rm A}/2+i(1+r_{\rm BM} e^{i\phi_{\rm B}})\Gamma_{\rm B}/2}
r_{\rm BM}e^{-2ik_{\rm B}z_{\rm M}},\\
r_{\rm BB}(\omega) &=&  \frac{\alpha+i\gamma_{\rm e}/2-i(1+e^{-i\phi_{\rm B}}/r_{\rm BM})\Gamma_{\rm B}/2+i(1+r_{\rm AM}e^{i\phi_{\rm A}})\Gamma_{\rm A}/2}
{\alpha+i\gamma_{\rm e}/2+i(1+r_{\rm AM} e^{i\phi_{\rm A}})\Gamma_{\rm A}/2+i(1+r_{\rm BM} e^{i\phi_{\rm B}})\Gamma_{\rm B}/2}
r_{\rm BM}e^{-2ik_{\rm B}z_{\rm M}},\label{eq:tBB_0}\\
c^{\rm (B)}_{\rm e}(\omega) &=& \frac{e^{-ik_{\rm B}z_0}(1+r_{\rm BM}e^{i\phi_{\rm B}})V_{\rm B}/\sqrt{2\pi}}
{\alpha+i\gamma_{\rm e}/2+i(1+r_{\rm AM} e^{i\phi_{\rm A}})\Gamma_{\rm A}/2+i(1+r_{\rm BM} e^{i\phi_{\rm B}})\Gamma_{\rm B}/2},\\
c^{\rm (B)}_{\rm s}(\omega) &=& \frac{e^{-ik_{\rm B}z_0}(1+r_{\rm BM}e^{i\phi_{\rm B}})V_{\rm B}\Omega/[2\sqrt{2\pi}(\omega-\omega_{\rm e}+\Delta_{\rm es})]}
{\alpha+i\gamma_{\rm e}/2+i(1+r_{\rm AM} e^{i\phi_{\rm A}})\Gamma_{\rm A}/2+i(1+r_{\rm BM} e^{i\phi_{\rm B}})\Gamma_{\rm B}/2}.
\end{eqnarray}
According to the Lippmann-Schwinger formalism, we can obtain the output state
\begin{eqnarray}
|\Psi^{(\rm B)}_{\rm out}\rangle 
&=& r_{\rm AB}(\omega) \int{\rm d}z \frac{e^{ik_{\rm A}z}}{\sqrt{2\pi}} a_{\rm R}^{\dagger}(z) |{\rm g},0\rangle
+r_{\rm BB}(\omega) \int{\rm d}z \frac{e^{ik_{\rm B}z}}{\sqrt{2\pi}} b_{\rm R}^{\dagger}(z) |{\rm g},0\rangle\nonumber\\
&=& r_{\rm AB}(\omega) |\Psi^{(\rm A)}_{\rm R}\rangle + r_{\rm BB}(\omega) |\Psi^{(\rm B)}_{\rm R}\rangle.
\end{eqnarray}

The output states $|\Psi^{(\rm A)}_{\rm out}\rangle$ and $|\Psi^{(\rm B)}_{\rm out}\rangle$ are determined by the coefficients $r_{\rm AA}(k)$, $r_{\rm BA}(k)$, $r_{\rm AB}(k)$ and $r_{\rm BB}(k)$.
Here we put the emitter at the center of waveguide corss-section $(x,y)=(a/2,b/2)$.
We set the waveguide cross-section size $a=b$, and we have $k_{\rm A}=k_{\rm B}$.
The reflection coefficients of the waveguide end are set as $r_{\rm AM}=r_{\rm BM}=-1$.
The separation $d$ between the emitter and the waveguide end is chosen to satisfy the condition $\phi_{\rm A}=\phi_{\rm B} \approx (2m+1)\pi$, where $m$ is an integer, for the frequency bands we are interested in. Under this condition, the emitter can interact with the symmetric superposition of the left- and right-propagating photon modes.
The phase factor $e^{-2ik_{\rm A}z_{\rm M}}$ and 
$e^{-2ik_{\rm B}z_{\rm M}}$ in these four expressions results from the photon's propagation in the waveguide, and its value does not affect the polarization conversion.
Here we can choose the coordinate $z_{\rm M}$ such that $-2k_{\rm A}z_{\rm M}=-2k_{\rm B}z_{\rm M}\approx (2n+1)\pi$, where $n$ is an integer.
Then from Eqs.~(\ref{eq:tAA_0}), (\ref{eq:tBA_0}), (\ref{eq:tAB_0}) and (\ref{eq:tBB_0}), we can obtain the expressions of the scattering matrix elements $r_{\rm AA}(k)$, $r_{\rm BA}(k)$, $r_{\rm AB}(k)$ and $r_{\rm BB}(k)$ in the main text.

\section{Obtaining arbitrary output state from an arbitrary input state}

The scattering matrix is
\begin{eqnarray}
{\mathcal{S}}=
\frac{1}{1-i\alpha/(2\Gamma_0)}
\left[
  \begin{array}{cc}
    -\cos 2\theta-i\alpha/(2\Gamma_0) & -\sin 2\theta \\
    -\sin 2\theta & \cos 2\theta-i\alpha/(2\Gamma_0) \\
  \end{array}
\right].
\end{eqnarray}
Consider an arbitrary input state $|\Psi_{\rm in}\rangle=[I_{\rm A} e^{i\xi_{\rm I}},I_{\rm B}]^{\rm T}$ and an arbitrary output state $|\Psi_{\rm out}\rangle=[O_{\rm A} e^{i(\xi_{\rm co}+\xi_{\rm O})},O_{\rm B}e^{i\xi_{\rm co}}]^{\rm T}$.
Here $0 \leq I_{\rm A}\leq 1$, $0 \leq I_{\rm B}\leq 1$, $0 \leq O_{\rm A}\leq 1$, $0 \leq O_{\rm B}\leq 1$, $I_{\rm A}^2 + I_{\rm B}^2 = 1$ and $O_{\rm A}^2 + O_{\rm B}^2 = 1$.
The phases $-\pi \leq \xi_{\rm I} < \pi$, $-\pi \leq \xi_{\rm O} < \pi$.
$\xi_{\rm co}$ is the common phase of the two components of the output state and is unknown. By expanding the equation $|\Psi_{\rm out}\rangle=\mathcal{S}|\Psi_{\rm in}\rangle$, we obtain the a set of equations
\begin{eqnarray}
O_{\rm A} e^{i(\xi_{\rm co}+\xi_{\rm O})} &=& \frac{-\cos 2\theta-i\alpha/(2\Gamma_0)}{1-i\alpha/(2\Gamma_0)} I_{\rm A} e^{i\xi_{\rm I}}
+ \frac{-\sin 2\theta}{1-i\alpha/(2\Gamma_0)} I_{\rm B},\\
O_{\rm B} e^{i\xi_{\rm co}} &=& \frac{-\sin 2\theta}{1-i\alpha/(2\Gamma_0)} I_{\rm A} e^{i\xi_{\rm I}}
+ \frac{\cos 2\theta-i\alpha/(2\Gamma_0)}{1-i\alpha/(2\Gamma_0)} I_{\rm B}.
\end{eqnarray}
Solving this set of equations, we obtain two sets of solutions which are given by
\begin{eqnarray}
\alpha^{(1)} &=& \frac{2\Gamma_0(I_{\rm A}I_{\rm B} \sin \xi_{\rm I} + O_{\rm A}O_{\rm B} \sin \xi_{\rm O})}
{\sqrt{(I_{\rm A}^2- O_{\rm A}^2)^2+(I_{\rm A}I_{\rm B} \cos \xi_{\rm I} - O_{\rm A}O_{\rm B} \cos \xi_{\rm O})^2}},\\
\sin 2\theta^{(1)} &=& \frac{I_{\rm A}^2- O_{\rm A}^2}
{\sqrt{(I_{\rm A}^2- O_{\rm A}^2)^2+(I_{\rm A}I_{\rm B} \cos \xi_{\rm I} - O_{\rm A}O_{\rm B} \cos \xi_{\rm O})^2}},\\
\cos 2\theta^{(1)} &=& \frac{-I_{\rm A}I_{\rm B} \cos \xi_{\rm I} + O_{\rm A}O_{\rm B} \cos \xi_{\rm O}}
{\sqrt{(I_{\rm A}^2- O_{\rm A}^2)^2+(I_{\rm A}I_{\rm B} \cos \xi_{\rm I} - O_{\rm A}O_{\rm B} \cos \xi_{\rm O})^2}},\\
e^{i\xi_{\rm co}^{(1)}} &=& \frac{-I_{\rm A}O_{\rm B}e^{i\xi_{\rm I}} + I_{\rm B}O_{\rm A}e^{-i\xi_{\rm O}}}
{\sqrt{(I_{\rm A}^2- O_{\rm A}^2)^2+(I_{\rm A}I_{\rm B} \cos \xi_{\rm I} - O_{\rm A}O_{\rm B} \cos \xi_{\rm O})^2} - i(I_{\rm A}I_{\rm B} \sin \xi_{\rm I} + O_{\rm A}O_{\rm B} \sin \xi_{\rm O})},
\end{eqnarray}
or
\begin{eqnarray}
\alpha^{(2)} &=& -\frac{2\Gamma_0(I_{\rm A}I_{\rm B} \sin \xi_{\rm I} + O_{\rm A}O_{\rm B} \sin \xi_{\rm O})}
{\sqrt{(I_{\rm A}^2- O_{\rm A}^2)^2+(I_{\rm A}I_{\rm B} \cos \xi_{\rm I} - O_{\rm A}O_{\rm B} \cos \xi_{\rm O})^2}},\\
\sin 2\theta^{(2)} &=& -\frac{I_{\rm A}^2- O_{\rm A}^2}
{\sqrt{(I_{\rm A}^2- O_{\rm A}^2)^2+(I_{\rm A}I_{\rm B} \cos \xi_{\rm I} - O_{\rm A}O_{\rm B} \cos \xi_{\rm O})^2}},\\
\cos 2\theta^{(2)} &=& -\frac{-I_{\rm A}I_{\rm B} \cos \xi_{\rm I} + O_{\rm A}O_{\rm B} \cos \xi_{\rm O}}
{\sqrt{(I_{\rm A}^2- O_{\rm A}^2)^2+(I_{\rm A}I_{\rm B} \cos \xi_{\rm I} - O_{\rm A}O_{\rm B} \cos \xi_{\rm O})^2}},\\
e^{i\xi_{\rm co}^{(2)}} &=& \frac{I_{\rm A}O_{\rm B}e^{i\xi_{\rm I}} - I_{\rm B}O_{\rm A}e^{-i\xi_{\rm O}}}
{\sqrt{(I_{\rm A}^2- O_{\rm A}^2)^2+(I_{\rm A}I_{\rm B} \cos \xi_{\rm I} - O_{\rm A}O_{\rm B} \cos \xi_{\rm O})^2} + i(I_{\rm A}I_{\rm B} \sin \xi_{\rm I} + O_{\rm A}O_{\rm B} \sin \xi_{\rm O})}.
\end{eqnarray}
With each set of solution, the system can transform the input state $|\Psi_{\rm in}\rangle$ into the output state $|\Psi_{\rm out}\rangle$. The difference between the two sets of solutions is $\alpha^{(1)}=-\alpha^{(2)}$, $\theta^{(1)}-\theta^{(2)}=\pm \pi/2$, and $e^{i\xi_{\rm co}^{(1)}}$ and $e^{i\xi_{\rm co}^{(2)}}$ are different phases in the output states, where we define $0 \leq \theta <\pi$.

\section{Rotation of a A-mode linearly polarized photon}
\begin{figure}[!ht]
\centering
\includegraphics[width=0.3\columnwidth]{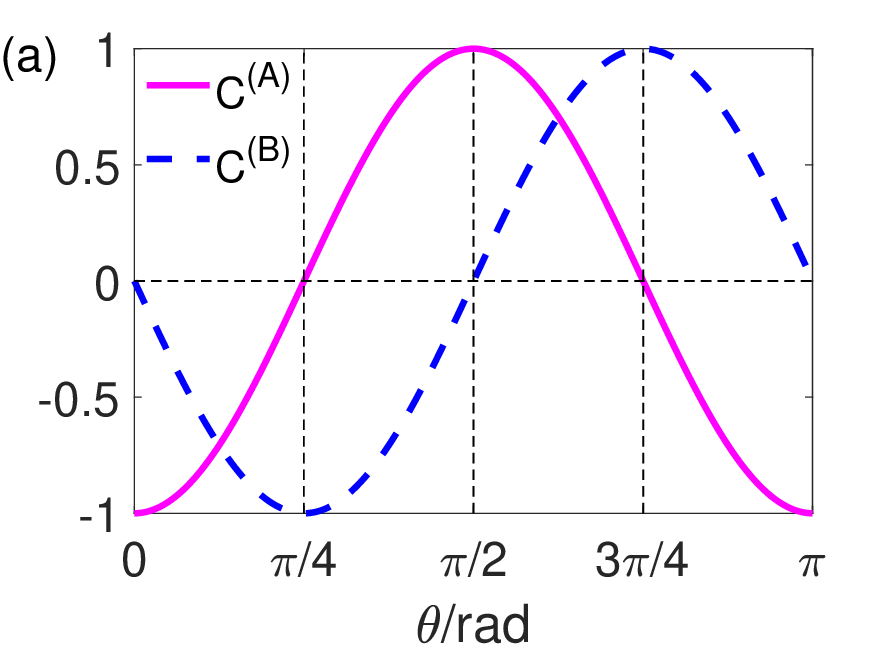}
\includegraphics[width=0.3\columnwidth]{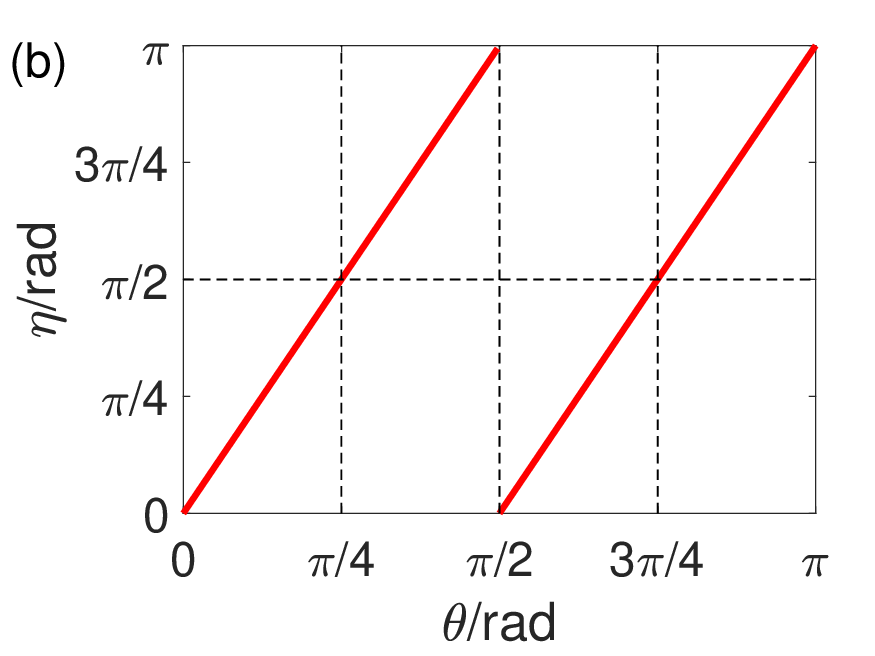}
\caption{The rotation of a horizontally linearly polarized input photon (A-mode phton).
(a) The coefficients of A mode $C^{\rm (A)}$ and B mode $C^{\rm (B)}$ in the output photon as a function of the electric dipole moment direction $\theta$ of the emitter.
(b) Polarization direction $\eta$ of the output photon at point $(x, y) = (a/2, a/2)$. Here $\eta$ is defined in the interval $0 \leq \eta < \pi$ and other parameters are $\alpha = 0$ and $\gamma_{\rm e} = 0$. }
\label{fig:Rotation}
\end{figure}

Considering the condition $\alpha=0$, i.e., $\Omega=2\sqrt{(\Delta_{\rm ge}-\Delta_{\rm es})\Delta_{\rm ge}}$, the scattering matrix is
\begin{eqnarray}
{\mathcal{S}_{\rm rot}}
=
\left[
  \begin{array}{cc}
    -\cos2\theta & -\sin2\theta \\
    -\sin2\theta & \cos2\theta \\
  \end{array}
\right].
\label{eq:S3}
\end{eqnarray}
When the incident photon is in the A mode (i.e., horizontal polarization, polarization degree $\eta=0$), the coefficients of A and B modes in the output state are $C^{\rm (A)}=-\cos2\theta$ and $C^{\rm (B)} = - \sin2\theta$, respectively, which are shown in Fig. \ref{fig:Rotation}(a). 
The polarization direction of the output photon is $\eta = 2\theta$ for $0 \leq \theta < \pi/2$ and $\eta = 2\theta - \pi$ for $\pi/2 \leq \theta < \pi$, which are shown in Fig. \ref{fig:Rotation}(b).

\section{Derivation of the Stokes parameters of the output photon when a horizontally polarized photon is input}
In the section ``Arbitrary polarization in the Poincar{\'e} sphere" of the main text, we consider a photon with horizontal polarization is input into the system. In the ideal situation with no dissipation $\gamma_2=0$, the output photon is 
\begin{eqnarray}
|\Psi_{\rm out}\rangle=r_{\rm AA}(k)|\Psi^{(\rm A)}_{\rm R}\rangle+r_{\rm BA}(k)|\Psi^{(\rm B)}_{\rm R}\rangle,
\end{eqnarray}
where
\begin{eqnarray}
r_{\rm AA}(k) &=& \frac{\sin^2\theta-\cos^2\theta-i\alpha/(2\Gamma_0)}{1-i\alpha/(2\Gamma_0)},\\
r_{\rm BA}(k) &=& \frac{-2\sin\theta\cos\theta}{1-i\alpha/(2\Gamma_0)}.
\end{eqnarray}
Here, $\theta$ is the direction of the emitter electric dipole moment, and  $\alpha=\Omega^2/[4(\Delta_{\rm ge}-\Delta_{\rm es})]-\Delta_{\rm ge}$.
From coefficients $r_{\rm AA}(k)$ and $r_{\rm BA}(k)$, we can obtain the Stokes parameters of the output photon~\cite{Born1999}
\begin{eqnarray}
s_1 &=& \frac{\cos4\theta+\alpha^2/(4\Gamma_0^2)}{1+\alpha^2/(4\Gamma_0^2)},\\
s_2 &=& \frac{\sin4\theta}{1+\alpha^2/(4\Gamma_0^2)},\\
s_3 &=& \frac{(\alpha/\Gamma_0)\sin2\theta}{1+\alpha^2/(4\Gamma_0^2)}.
\end{eqnarray}
Point $\vec{s}=(s_1,\ s_2,\ s_3)$ is the point ${\rm P}(\theta,\alpha)$ in Fig. 3(a) in the main text. For a given $\theta$, with $\alpha$ changes from $-\infty$ to $\infty$, point ${\rm P}(\theta,\alpha)$ moves on the yellow circle along the direction of the red arrows on the Poincar{\'e} sphere. Then with $\theta$ changes from $0$ to $\pi/2$, the circle can scan over the whole Poincar{\'e} sphere. Thus, we can obtain an output photon with arbitrary polarization.

\section{Tunable working frequency of our scheme}

\begin{figure}[!ht]
\centering
\includegraphics[width=0.3\columnwidth]{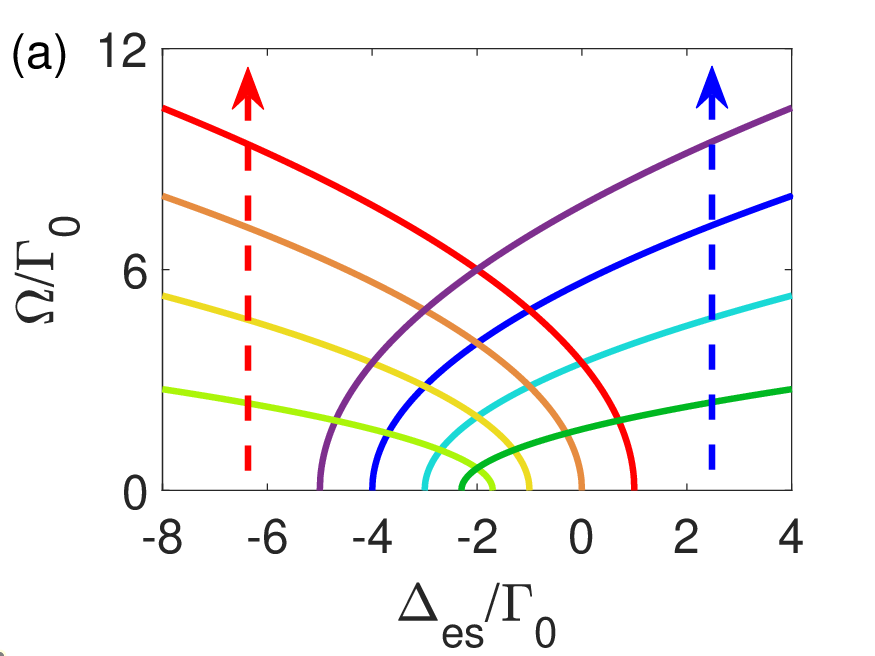}
\includegraphics[width=0.3\columnwidth]{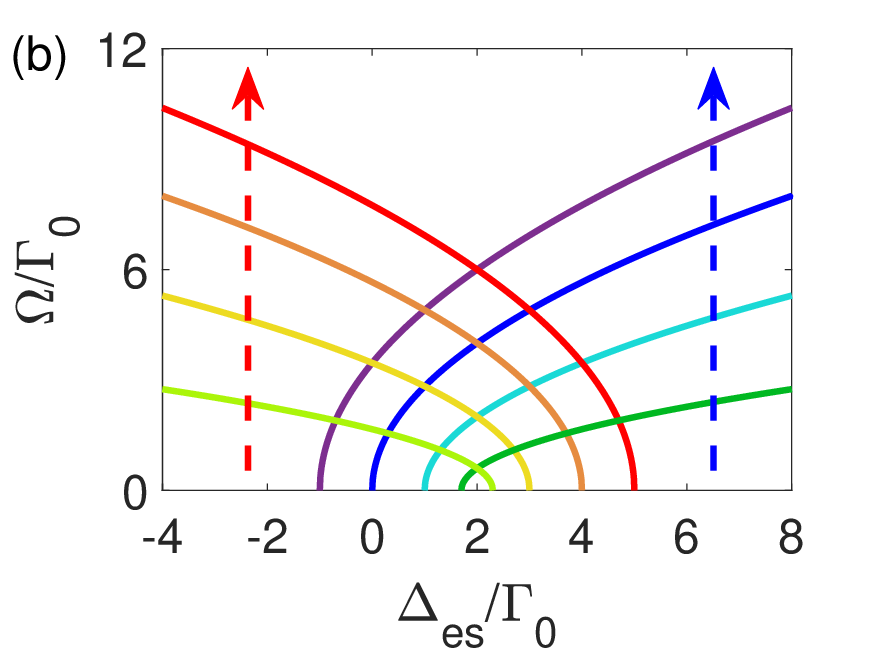}
\includegraphics[width=0.3\columnwidth]{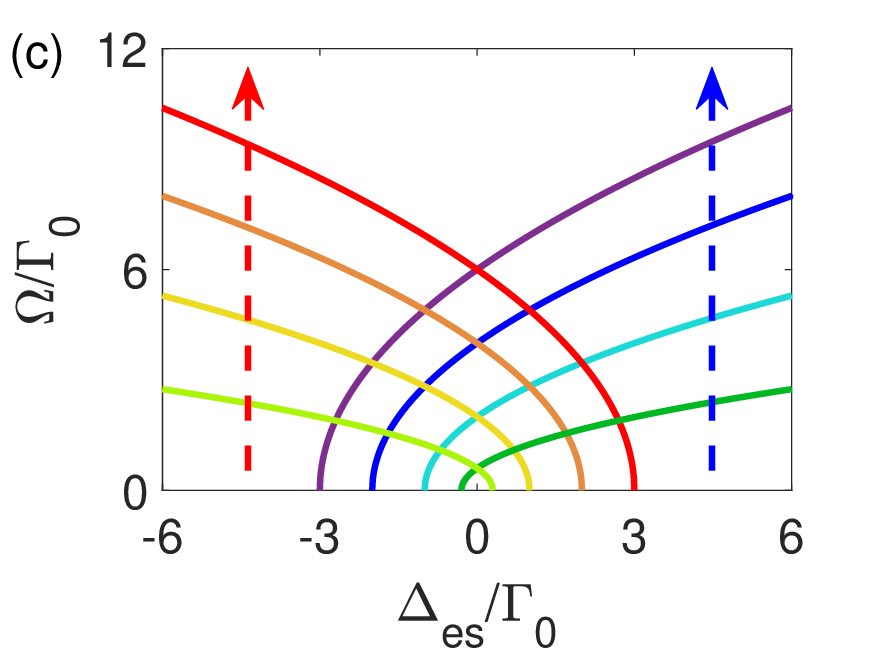}
\caption{(a) Coupling strength $\Omega$ as functions of detuning $\Delta_{\rm es}$ according to the condition $\Omega=2\sqrt{(\Delta_{\rm ge}-\Delta_{\rm es})(\Delta_{\rm ge}+2\Gamma_0)}$.
The values of $\Delta_{\rm ge}/\Gamma_0$ for different curves are, along the red arrow, -1.7, -1, 0, 1, and along the blue arrow, -2.3, -3, -4, -5, respectively.
(b) Coupling strength $\Omega$ as functions of detuning $\Delta_{\rm es}$ according the condition $\Omega=2\sqrt{(\Delta_{\rm ge}-\Delta_{\rm es})(\Delta_{\rm ge}-2\Gamma_0)}$.
The values of $\Delta_{\rm ge}/\Gamma_0$ for different curves are, along the red arrow, 2.3, 3, 4, 5, and along the blue arrow, 1.7, 1, 0, -1, respectively.
(c) Coupling strength $\Omega$ as functions of detuning $\Delta_{\rm es}$ according the condition $\Omega=2\sqrt{(\Delta_{\rm ge}-\Delta_{\rm es})\Delta_{\rm ge}}$.
The values of $\Delta_{\rm ge}/\Gamma_0$ for different curves are, along the red arrow, 0.3, 1, 2, 3, and along the blue arrow, -0.3, -1, -2, -3, respectively.
\label{fig:pianzhen_two_alpha_theta_sup}}
\end{figure}

In our scheme, the matrix elements $r_{\rm AA}$, $r_{\rm BA}$, $r_{\rm AB}$, and $r_{\rm BB}$ of the scattering matrix $\mathcal{S}$ can be determined by the emission rates $\Gamma_{\rm A}$ , $\Gamma_{\rm B}$, and the parameter $\alpha=\Omega^2/[4(\Delta_{\rm ge}-\Delta_{\rm es})]-\Delta_{\rm ge}$ in the ideal situation with no dissipation $\gamma_{\rm e}=0$.
When we choose an isotropic emitter or an anisotropic emitter with electric dipole moment direction $\theta=\pi/4$, we have $\Gamma_{\rm A}=\Gamma_{\rm B}=\Gamma_0$. 
Then if the condition $\alpha=2\Gamma_0$, i.e., $\Omega=2\sqrt{(\Delta_{\rm ge}-\Delta_{\rm es})(\Delta_{\rm ge}+2\Gamma_0)}$, is satisfied, the scattering matrix is given by $\mathcal{S}_+$ in Eq.~(6) in the main text, with which the polarization conversion $|{\rm H}\rangle \leftrightarrow |{\rm R}\rangle$ and $|{\rm V}\rangle \leftrightarrow |{\rm L}\rangle$ can be realized.
We can see that for an arbitrary given input photon frequency detuning $\Delta_{\rm ge}$, 
we can always find proper coupling strength $\Omega$ and frequency detuning $\Delta_{\rm es}$ of the control field to satisfy the condition, which means the working frequency of our scheme is tunable. In addition, $\Omega$ is a function of $\Delta_{\rm es}$, which means there are infinite number of ($\Delta_{\rm es}$, $\Omega$) satisfying the condition, and we can choose some practical ($\Delta_{\rm es}$, $\Omega$) when our scheme is used to practical systems.
In Fig.~\ref{fig:pianzhen_two_alpha_theta_sup}(a), we plot $\Omega$ as functions of $\Delta_{\rm es}$ according to the condition $\Omega=2\sqrt{(\Delta_{\rm ge}-\Delta_{\rm es})(\Delta_{\rm ge}+2\Gamma_0)}$ for several different values of $\Delta_{\rm ge}$.
Similarly, when $\Gamma_{\rm A}=\Gamma_{\rm B}=\Gamma_0$ and $\alpha=-2\Gamma_0$, i.e., $\Omega=2\sqrt{(\Delta_{\rm ge}-\Delta_{\rm es})(\Delta_{\rm ge}-2\Gamma_0)}$, the scattering matrix is given by $\mathcal{S}_-$ in Eq.~(6) in the main text, with which the polarization conversion $|{\rm H}\rangle \leftrightarrow |{\rm L}\rangle$ and $|{\rm V}\rangle \leftrightarrow |{\rm R}\rangle$ can be realized. In Fig.~\ref{fig:pianzhen_two_alpha_theta_sup}(b), we plot $\Omega$ as functions of $\Delta_{\rm es}$ for several different values of $\Delta_{\rm ge}$.
When $\alpha=0$, i.e., $\Omega=2\sqrt{(\Delta_{\rm ge}-\Delta_{\rm es})\Delta_{\rm ge}}$, the scattering matrix is given by $\mathcal{S}_{\rm Rot}$ in Eq.~(10) in the main text, with which the polarization rotation of linearly polarized photon can be realized. 
In Fig.~\ref{fig:pianzhen_two_alpha_theta_sup}(c), we plot $\Omega$ as functions of $\Delta_{\rm es}$ for several different values of $\Delta_{\rm ge}$ according to this condition.

\section{Discussions about the effects of nonideal conditions}
\subsection{The impact of a slight difference between the height and the width of the waveguide on the scheme's effectiveness}

\begin{figure}[!ht]
\centering
\includegraphics[width=0.32\columnwidth]{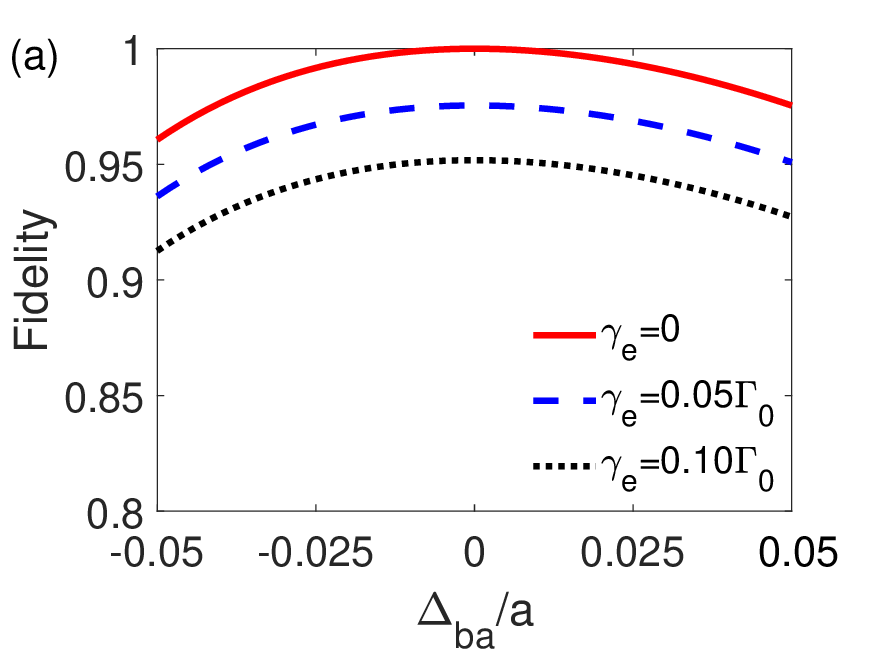}
\includegraphics[width=0.32\columnwidth]{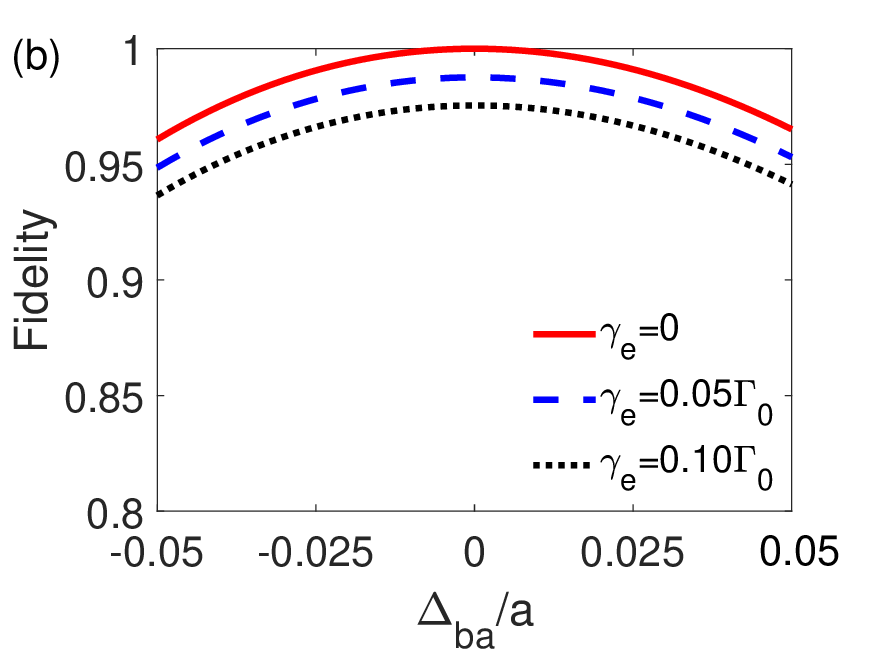}
\includegraphics[width=0.32\columnwidth]{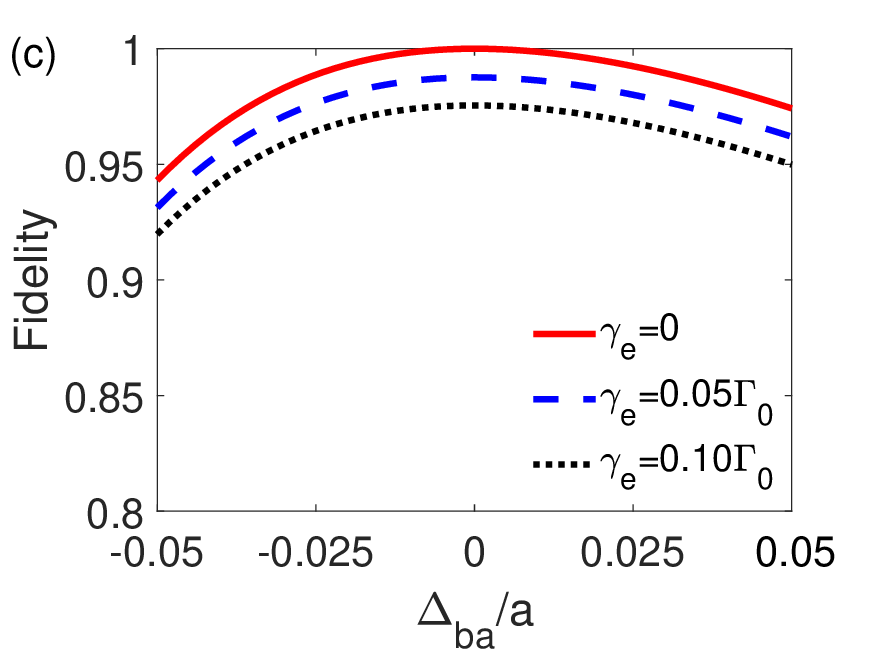}
\caption{The fidelity of three polarization conversions as functions of the difference $\Delta_{ba}=b-a$ between the height $b$ and the width $a$ of the waveguide with different values of the emitter dissipation $\gamma_{\rm e}$. (a) The conversion of a horizontally linearly polarized photon to a vertically linearly polarized photon ($|{\rm H}\rangle \to |{\rm V}\rangle$). (b) The conversion of a vertically linearly polarized photon to a left-handed circularly polarized photon ($|{\rm V}\rangle \to |{\rm L}\rangle$). (c) The conversion of a right-handed circularly polarized photon to a vertically linearly polarized photon ($|{\rm R}\rangle \to |{\rm V}\rangle$). In all the subfigures, other parameters are $(x,y)=(a/2,\ b/2)$, $d=0.75\lambda_{{\rm B}z}=1.806\ a$, $r_{\rm M}=-1$. Here $\lambda_{{\rm B}z}=2\pi/k_{{\rm B}z}$, and $k_{{\rm B}z}=\sqrt{k^2-(\pi/a)^2}$ is the $z$-direction wave vector of mode B. Here we set $k=1.3\pi/a$.
\label{fig:Nonideal_abnotequal}}
\end{figure}
In a more realistic scenario, the waveguides height and width may differ slightly. Here, we consider the situation that there is a 0{\%}-5{\%} difference between the width $a$ and the height $b$, and find that the fidelity of the polarization conversion can be still very high. 
We study the influence of imperfect dimension on three different polarization conversions: (i) The conversion of a horizontally linearly polarized input photon $|{\rm H}\rangle$ to a vertically linearly polarized output photon $|{\rm V}\rangle$ [Fig.~\ref{fig:Nonideal_abnotequal}(a)]. (ii) The conversion of a vertically linearly polarized photon $|{\rm V}\rangle$ to a left-handed circularly polarized photon $|{\rm L}\rangle$[Fig.~\ref{fig:Nonideal_abnotequal}(b)]. (iii) The conversion of a right-handed circularly polarized photon $|{\rm R}\rangle$ to a vertically linearly polarized photon $|{\rm V}\rangle$ [Fig.~\ref{fig:Nonideal_abnotequal}(c)].
From the figures, we can see that the fidelity decreases slightly as $|\Delta_{ba}|$ increases. 
The reason is that the different $a$ and $b$ leads to the lift of the degeneracy between mode A and mode B, i. e., their wavelengths differ from each other. Considering that the mode A and mode B are standing waves in the semi-infinite waveguide, they will have different light intensities at the emitter’s position, which results in the different coupling strengths between the emitter and the two waveguide modes. The output photon state therefore slightly deviates from the ideal state, which leads to the decrease of the fidelity.
However, despite that the fidelity decreases as $|\Delta_{ab}|$ increases, the conversion fidelities can be still larger than 94{\%} for all the three cases when $|\Delta_{ab}|\leq 0.05a$ without external dissipation (i.e., $\gamma_{\rm e}=0$). Even if there is external dissipation (e.g., $\gamma_{\rm e}=0.1\Gamma_0$), the conversion efficiency can still be larger than 90{\%} for all the three cases when $|\Delta_{ab}| \leq 0.05a$. Under current fabrication technology, it is not difficult for $|\Delta_{ab}|$ to be controlled within 5{\%} and therefore the conversion fidelity can be larger than 90{\%} even if there is external dissipation $\gamma_{\rm e}=0.1\Gamma_0$.

\subsection{The influence of the quantum emitter's position within the x, y plane on the conversion efficiency and the overall viability of the scheme}

\begin{figure}[!ht]
\centering
\includegraphics[width=0.32\columnwidth]{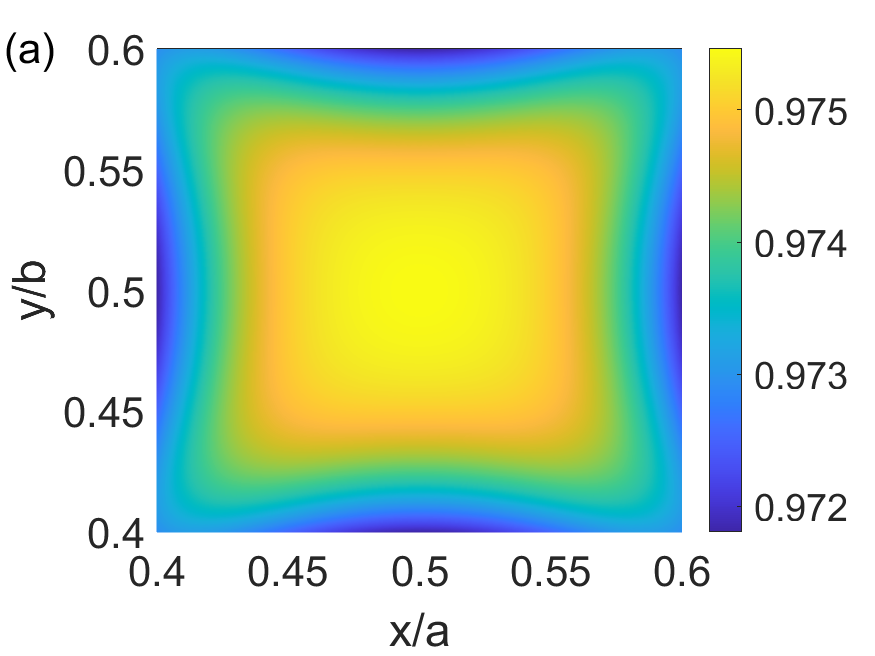}
\includegraphics[width=0.32\columnwidth]{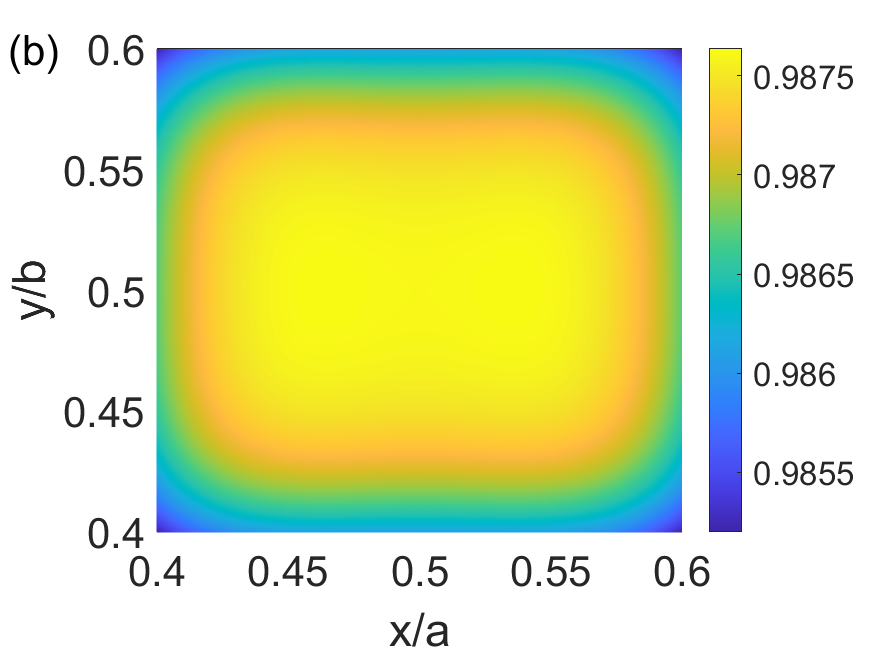}
\includegraphics[width=0.32\columnwidth]{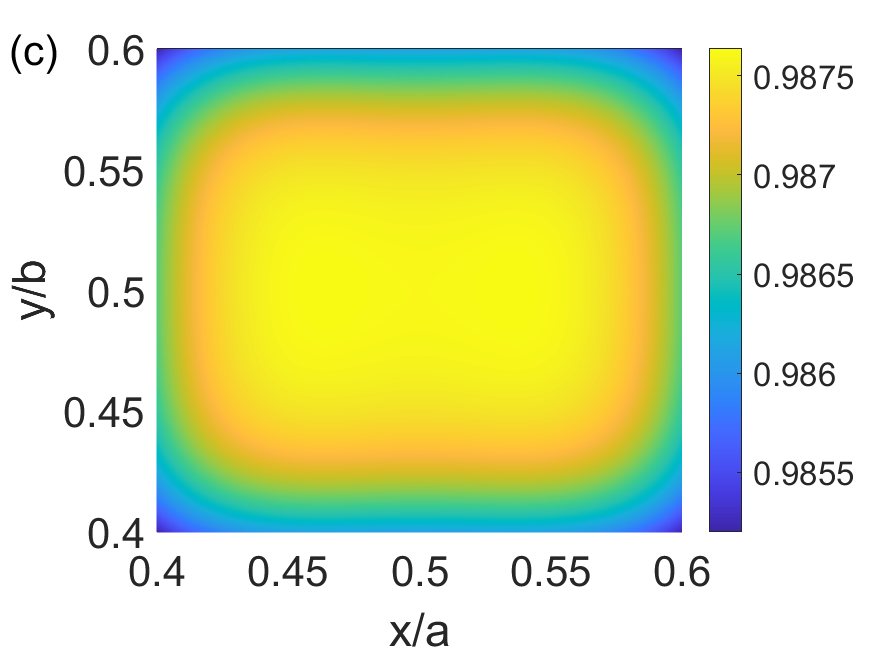}
\caption{The fidelity of three polarization conversions as functions of the emitter's position within the $x$, $y$ plane. (a) The conversion of a horizontally linearly polarized photon to  a vertically linearly polarized photon ($|{\rm H}\rangle \to |{\rm V}\rangle$). (b) The conversion of a vertically polarized photon to a left-handed circularly polarized photon ($|{\rm V}\rangle \to |{\rm L}\rangle$). (c) The conversion of a right-handed circularly polarized photon to a vertically polarized photon ($|{\rm R}\rangle \to |{\rm V}\rangle$).
In all the subfigures, other parameters are $\gamma_{\rm e}=0.05\Gamma_0$, $a=b$, $d=0.75\lambda_{{\rm B}z}=1.806\ a$, $r_{\rm M}=-1$. Here $\lambda_{{\rm B}z}=2\pi/k_{{\rm B}z}$, and $k_{{\rm B}z}=\sqrt{k^2-(\pi/a)^2}$ is the $z$-direction wave vector of mode B. Here we set $k=1.3\pi/a$.
\label{fig:Nonideal_xy}}
\end{figure}
We consider the influence of the quantum emitter’s position within the $x$, $y$ plane on the conversion efficiency of the scheme. The results are shown in Fig.~\ref{fig:Nonideal_xy}. Overall, if the emitter’s position deviates a bit from the center of the waveguide, the polarization conversion efficiency is not affected seriously. 
More specifically, we vary the emitter position with $0.4a\leq x \leq 0.6a$,  $0.4b\leq y \leq 0.6b$, and study the fidelities of three different polarization conversions: (i) The conversion of a horizontally linearly polarized input photon $|{\rm H}\rangle$ to a vertically linearly polarized output photon $|{\rm V}\rangle$ [Fig.~\ref{fig:Nonideal_xy}(a)]. (ii) The conversion of a vertically linearly polarized photon $|{\rm V}\rangle$ to a left-handed circularly polarized photon $|{\rm L}\rangle$ [Fig.~\ref{fig:Nonideal_xy}(b)]. (iii) The conversion of a right- handed circularly polarized photon $|{\rm R}\rangle$ to a vertically linearly polarized photon $|{\rm V}\rangle$ [Fig.~\ref{fig:Nonideal_xy}(c)]. For all three cases, we can see even if the position of the emitter off the center is high up to 25{\%}, the conversion efficiencies can all be greater than 97{\%}. For other conversions, we have the similar results which are shown here. Thus, our scheme is robust again the transverse position variations.

\subsection{The necessity of placing the emitter at the antinode position}

\begin{figure}[!ht]
\centering
\includegraphics[width=0.3\columnwidth]{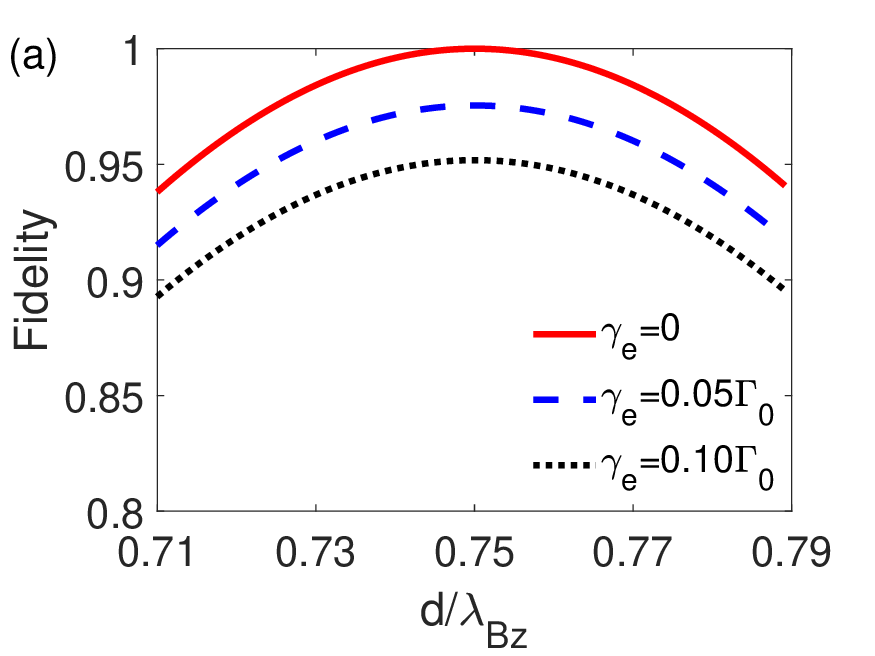}
\includegraphics[width=0.3\columnwidth]{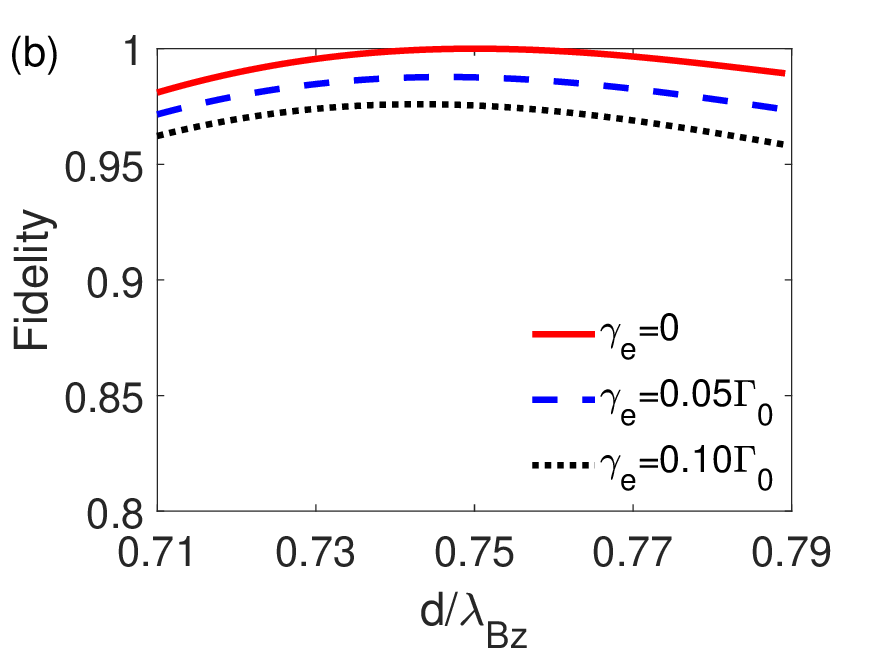}
\includegraphics[width=0.3\columnwidth]{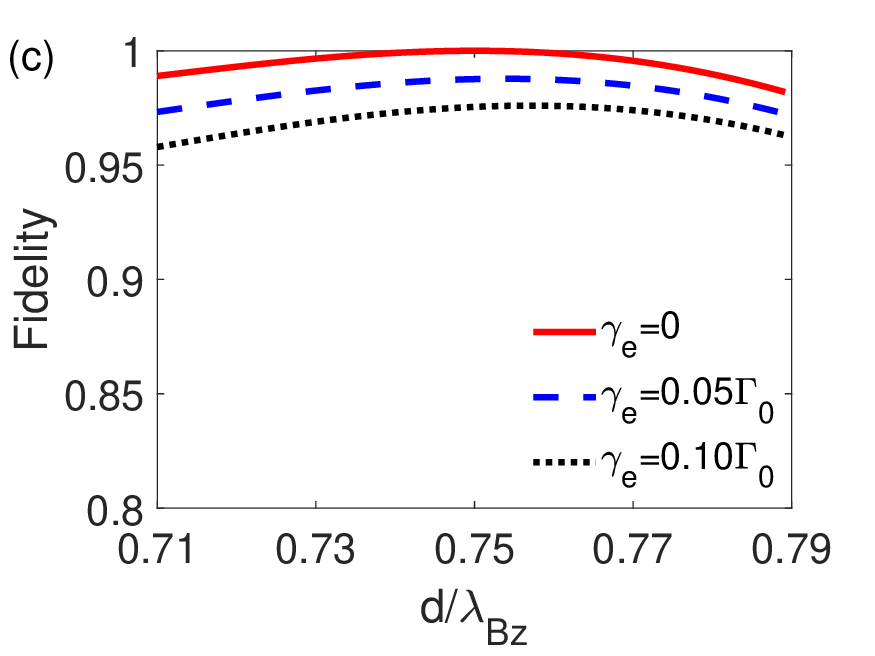}
\caption{The fidelity of three polarization conversions as functions of the separation $d$ between the emitter and the waveguide end with different values of the emitter dissipation $\gamma_{\rm e}$. (a) The conversion of a horizontally linearly polarized photon to a vertically linearly polarized photon ($|{\rm H}\rangle \to |{\rm V}\rangle$). (b) The conversion of a vertically polarized photon to a left-handed circularly polarized photon ($|{\rm V}\rangle \to |{\rm L}\rangle$). (c) The conversion of a right-handed circularly polarized photon to a vertically polarized photon ($|{\rm R}\rangle \to |{\rm V}\rangle$). 
Here $\lambda_{{\rm B}z}=2\pi/k_{{\rm B}z}$, and $k_{{\rm B}z}=\sqrt{k^2-(\pi/a)^2}$ is the $z$-direction wave vector of mode B. 
Here we set $k=1.3\pi/a$.
In all the subfigures, other parameters are $a=b$, $(x,y)=(a/2,\ b/2)$, $r_{\rm M}=-1$.
\label{fig:Nonideal_NotAntinode}}
\end{figure}
The emitter may not be exactly at the antinode position in a realistic scenario and it may affect the conversion efficiency. Here, we consider how the photon polarization conversion fidelity changes when the emitter is around the antinode position. Here we also consider three different polarization conversions (i.e., $|{\rm H}\rangle \to |{\rm V}\rangle$, $|{\rm V}\rangle \to |{\rm L}\rangle$, and $|{\rm R}\rangle \to |{\rm V}\rangle$) and the results are shown in Fig.~\ref{fig:Nonideal_NotAntinode}. From the results we can see that in all the three cases, with the ideal value $d=0.75\lambda_{{\rm B}z}=1.806\ a$, i.e., the emitter is at an antinode, the fidelity has maximum values, and when $d$ deviates from the ideal value, the fidelity decreases.
This is because at the antinode, light intensities of mode A and mode B have maximum values, and the coupling strengths between the emitter and the two modes have maximum values. The conversion between the two modes can happen with the ideal probability, and the fidelity has maximum values.
On the contrary, when the emitter position deviates from the antinode, the light intensities decrease, and the coupling strengths decrease. The conversion probability between modes A and B deviates from the ideal value, and the output photon state deviates from the ideal state.
However, despite that the fidelities decreases as $d$ deviates from the ideal value, the conversion fidelities in all three cases can still be larger than 0.95 when the deviation is within 5{\%} (i.e., $0.7125\lambda_{{\rm B}z}\leq d \leq 0.7875\lambda_{{\rm B}z}$) without external dissipation (i.e., $\gamma_{\rm e}=0$). Even if there is external dissipation, e.g., $\gamma_{\rm e}=0.1\Gamma_0$, the conversion fidelities for all three cases can be larger than 0.9 when the deviation is within 5{\%}. Thus, our scheme is also robust against the variation of the distance $d$ between the emitter and the mirror.

\subsection{The necessity of 100{\%} reflection at the waveguides end}

\begin{figure}[!ht]
\centering
\includegraphics[width=0.3\columnwidth]{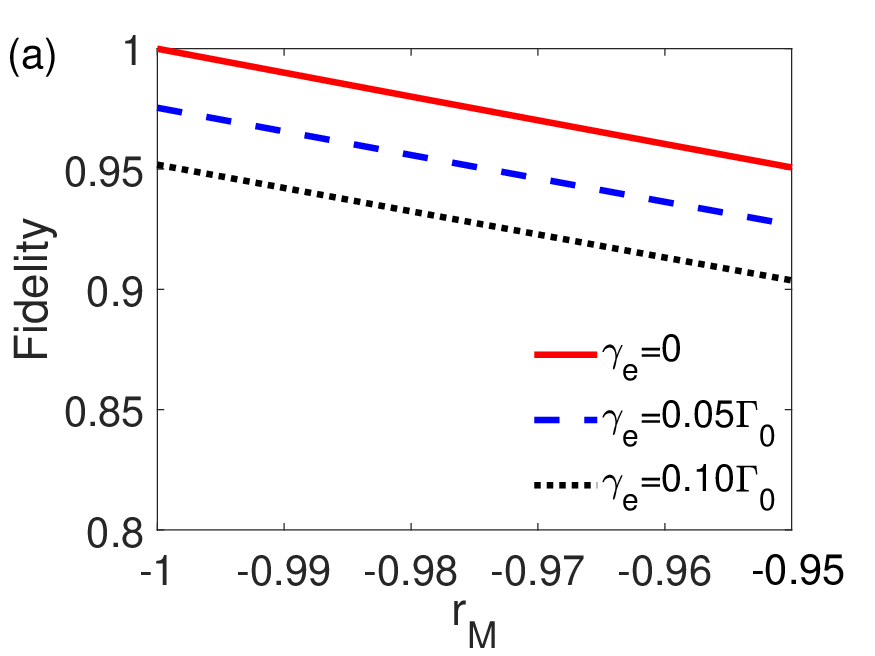}
\includegraphics[width=0.3\columnwidth]{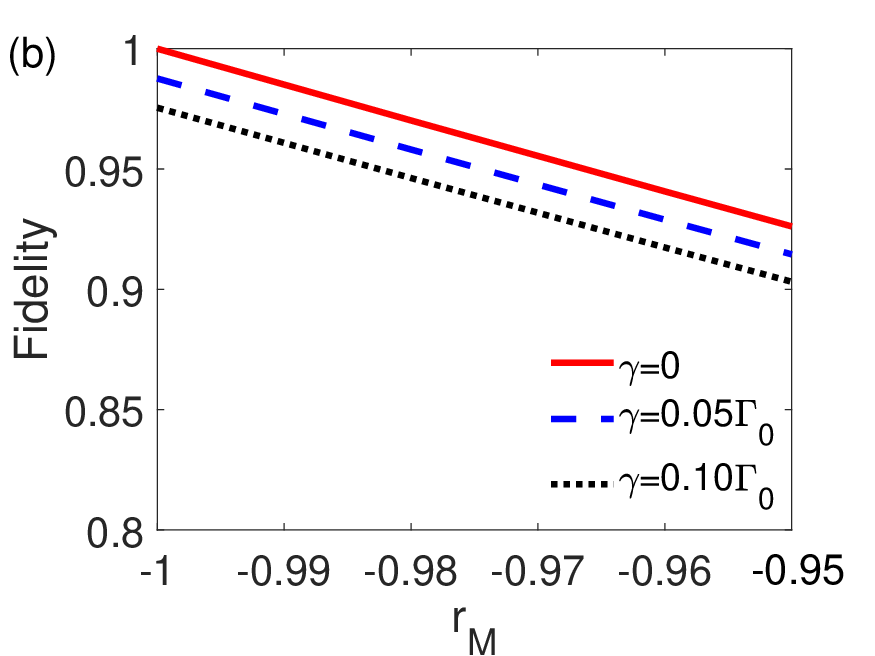}
\includegraphics[width=0.3\columnwidth]{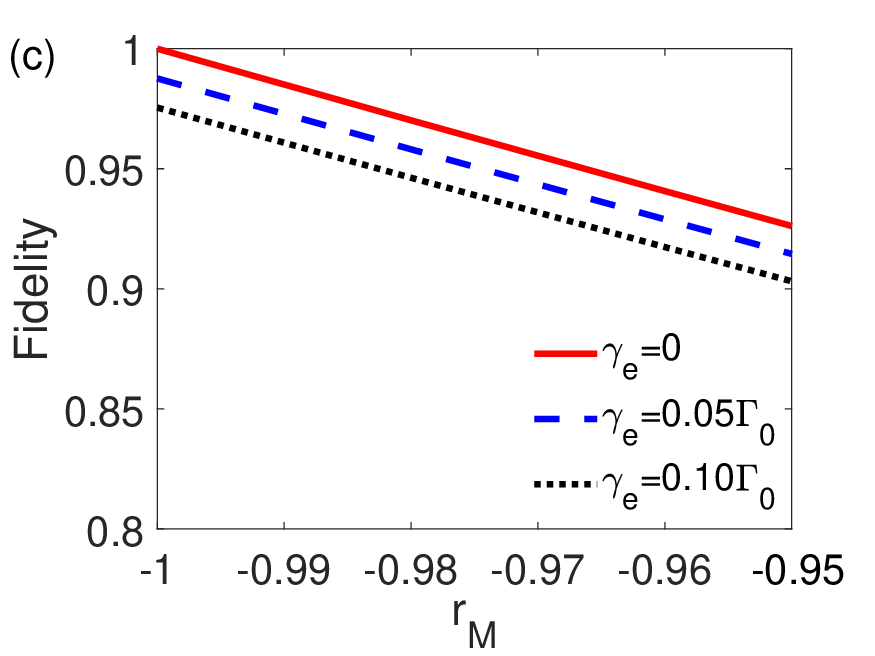}
\caption{The fidelity of three polarization conversions as functions of the waveguide end reflection $r_{\rm M}$ with different values of the emitter dissipation $\gamma_{\rm e}$. (a) The conversion of a horizontally linearly polarized photon to a vertically linearly polarized photon ($|{\rm H}\rangle \to |{\rm V}\rangle$). (b) The conversion of a vertically polarized photon to a left-handed circularly polarized photon ($|{\rm V}\rangle \to |{\rm L}\rangle$). (c) The conversion of a right-handed circularly polarized photon to a vertically polarized photon ($|{\rm R}\rangle \to |{\rm V}\rangle$). In all the subfigures, other parameters are $a=b$, $(x,y)=(a/2,\ b/2)$, $d=0.75\lambda_{{\rm B}z}=1.806\ a$. Here $\lambda_{{\rm B}z}=2\pi/k_{{\rm B}z}$, and $k_{{\rm B}z}=\sqrt{k^2-(\pi/a)^2}$ is the $z$-direction wave vector of mode B. Here we set $k=1.3\pi/a$.
\label{fig:Nonideal_ReflectionNot100}}
\end{figure}
In a more realistic scenario, the waveguide end reflectivity may not be exactly 100{\%}. Here, we study the polarization conversion fidelity for the three cases (i.e., $|{\rm H}\rangle \to |{\rm V}\rangle$, $|{\rm V}\rangle \to |{\rm L}\rangle$, and $|{\rm R}\rangle \to |{\rm V}\rangle$) when the reflectivity coefficient $r_{\rm M}$ varies from -1 to -0.95. The results are shown in Fig.~\ref{fig:Nonideal_ReflectionNot100} from which we can see that when the reflection coefficient $r_{\rm M}$ is -1, the fidelity has maximum value, and with $|r_{\rm M}|$ decreases, the fidelity decreases.
One reason is that the decrease of the reflection leads to the loss of the photon's energy and therefore the decrease of the probability of generating an output photon, which makes the fidelity decrease. 
Another reason is that the imperfect reflection makes reflected light is weaker than the incident light, and the standing wave is broken. 
In this situation, the wave can be seen as a superposition of a standing wave and a travelling wave. The standing wave can be converted into target polarized state completely, but the travelling wave can not be converted into target polarized state completely. 
As a result, the output states deviates form the ideal states, and the fidelity decreases.
However, although the fidelities in all three cases decrease as the reflectivity decreases, all of them are still larger than 90{\%}.

\subsection{Effect of the dissipation}
\begin{figure}[!ht]
\centering
\includegraphics[width=0.293\columnwidth]{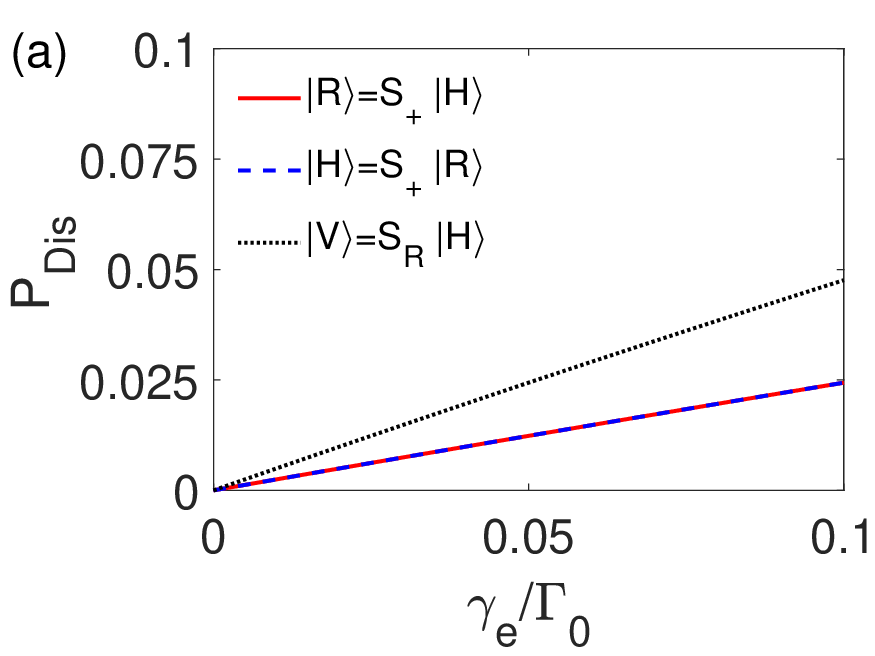}
\includegraphics[width=0.345\columnwidth]{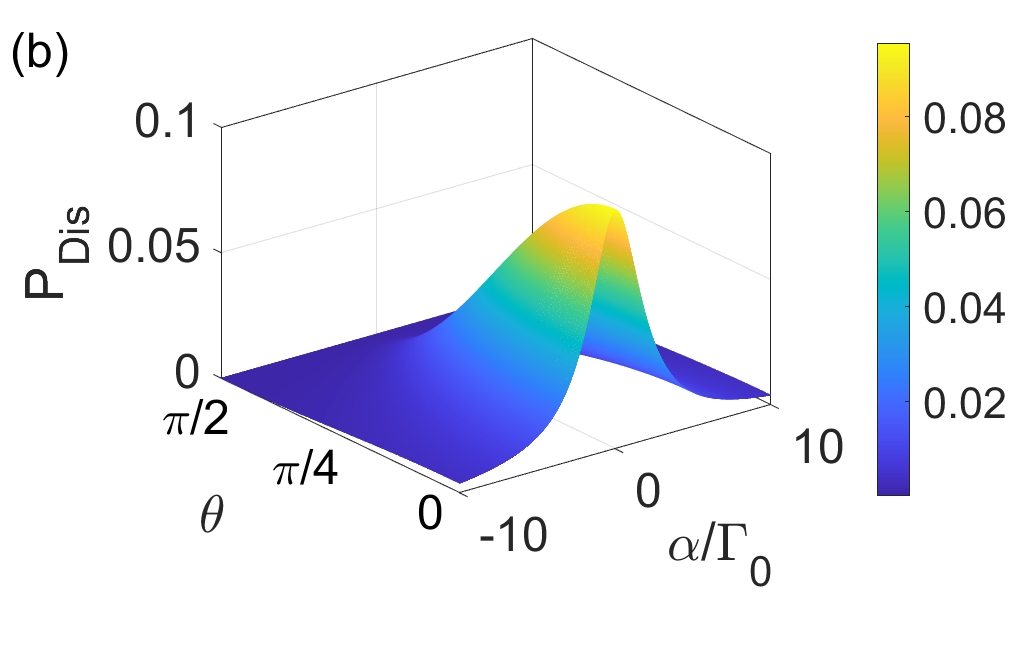}
\caption{(a) Dissipation probability $P_{\rm Dis}$ of the photon as functions of external dissipation rate $\gamma_e$. The three curves represent three different situations of polarization conversion as denoted in the legend. (b) Dissipation probability $P_{\rm Dis}$ as a function of $\alpha$ and $\theta$ when the input photon is horizontally polarized with $\gamma_{\rm e}=0.1\Gamma_0$.
\label{fig:dissipation}}
\end{figure}
In the above discussions the external dissipation $\gamma_{\rm e}$ is neglected and the polarization conversion efficiency can be 100\%.  Here, we consider the effect of the external dissipation. We first consider polarization conversions for three situations with $\gamma_{\rm e}/\Gamma_0\leq 0.1$: $|\rm R\rangle={\mathcal{S}_+}|\rm H\rangle$, $|\rm H\rangle={\mathcal{S}_+}|\rm R\rangle$ and $|\rm V\rangle={\mathcal{S}_{\rm rot}}|\rm H\rangle$.
With the increase of the external dissipation $\gamma_{\rm e}$, the dissipation rate $P_{\rm Dis}$ increases (Fig.~\ref{fig:dissipation}(a)).
For the the conversions $|\rm R\rangle={\mathcal{S}_+}|\rm H\rangle$ and $|\rm H\rangle={\mathcal{S}_+}|\rm R\rangle$, $P_{\rm Dis}<0.025$ (i.e., efficiency $>$ 97.5\%) for the whole range. 
For the conversion $|\rm V\rangle={\mathcal{S}_{\rm rot}}|\rm H\rangle$, $P_{\rm Dis}<0.05$ (i.e., efficiency $>$ 95\%). The low dissipation rate results from similar effect of EIT where large $\alpha$ can suppress the effect of $\gamma_{e}$ as shown in Fig.~\ref{fig:dissipation}(b).
We then consider a horizontally polarized input photon being scattered with $-\infty<\alpha<+\infty$ and $0\leq \theta \leq \pi/2$, and the polarization of the output photon can arrive all the points on the Poincar{\'e} sphere. 
For the whole value range of parameters, we have dissipation $P_{\rm Dis}<0.1$ (i.e., efficiency $>$ 90\%) for $\gamma_{\rm e}= 0.1\Gamma_0$. 
Thus, if the external dissipation rate is not very large which is currently experimentally achievable \cite{Scarpelli2019}, the conversion efficiency can still be larger than 90{\%} in our scheme.

\end{widetext}


\providecommand{\noopsort}[1]{}\providecommand{\singleletter}[1]{#1}%

\end{document}